\documentclass[journal,12pt,draftclsnofoot,onecolumn]{IEEEtran}
\usepackage{cite}
\usepackage{graphicx}
\usepackage{subfig}
\usepackage{booktabs}
\usepackage{tabularx}
\usepackage{multirow}
\usepackage{fancyhdr}
\usepackage[cmex10]{amsmath}
\usepackage{algorithm}
\usepackage{algpseudocode}
\usepackage{algorithmicx}
\usepackage{filecontents}
\usepackage[dvipsnames]{xcolor}
\usepackage[percent]{overpic}
\usepackage{todonotes}
\usepackage{enumitem}
\usepackage{arydshln}
\usepackage{verbatim}
\usepackage{soul} 
\usepackage{mathrsfs}

\begin{document}

\title{Channel Characterization of Diffusion-based Molecular Communication with Multiple Fully-absorbing Receivers}

\author{\IEEEauthorblockN{Marco~Ferrari\IEEEauthorrefmark{1}, Fardad Vakilipoor\IEEEauthorrefmark{2},
Eric~Regonesi\IEEEauthorrefmark{2},
Mariangela~Rapisarda\IEEEauthorrefmark{2}, Maurizio~Magarini\IEEEauthorrefmark{2}}\\ \IEEEauthorblockA{\IEEEauthorrefmark{1}CNR - IEIIT, DEIB, Politecnico di Milano, 20133 Milano, Italy\\
\IEEEauthorrefmark{2}Dipartimento di Elettronica, Informazione e Bioingegneria, Politecnico di Milano, 20133 Milano, Italy
}
}

\maketitle

\begin{abstract}
In this paper an analytical model is introduced to describe the impulse response of the diffusive channel between a pointwise transmitter and a given fully-absorbing (FA) receiver in a molecular communication (MC) system. The presence of neighbouring FA nanomachines in the environment is taken into account by describing them as sources of negative molecules. 
\textcolor{black}{The channel impulse responses of all the receivers are linked in a system of integral equations. The solution of the system with two receivers is obtained analytically. For a higher number of receivers the system of integral equations is solved numerically.}
It is also shown that the channel impulse response shape is distorted by the presence of the interferers. For instance, there is a time shift of the peak in the number of absorbed molecules compared to the case without interference, as predicted by the proposed model. The analytical derivations are validated by means of particle based simulations.
\end{abstract}
\begin{IEEEkeywords}
Molecular communication, channel modelling, diffusion, multiple receivers, interference. 
\end{IEEEkeywords}

\section{Introduction} \label{intro}

The recent advancements in synthetic biology and bio-nanotechnology have opened the way to new opportunities in many application fields, such as medicine, tissues and materials engineering, and environmental monitoring and preservation~\cite{akyildiz2019moving,farsad2016comprehensive}. The basis of such progresses is the possibility that bio-nanomachines cooperate, which is possible only when they are organized in a network and a communication can be established among the constituting nodes~\cite{akyildiz2019information}. Bio-nanomachines in a fluid environment can communicate, for instance, through an exchange of signals or molecules, \textit{i.e.}, through molecular communication (MC)~\cite{nakano2013molecular}.

Despite the initial studies of MC have focused on static situations, where transmitting and receiving bio-nanomachines are in fixed positions, some recent research works have relaxed this condition to move towards more realistic dynamic scenarios with mobile nodes. 
The paper by Nakano \textit{et al.}~\cite{nakano2019methods} offers a comprehensive overview of the progresses in the emerging research area of mobile MC. A fundamental characteristic of mobile MC is that groups of bio-nanomachines collaborate to provide functionalities, thus overcoming the problem of limited computational capabilities of a single bio-nanomachine.




Mobility leads to an improvement of communication performance such as, for example, in active transport based on molecular motors~\cite{keya2018dna} or in MC via diffusion (MCvD)~\cite{yilmaz2014three}. Also, mobility allows for the implementation of network functionalities as discussed, for instance, in~\cite{ishiyama2018epidemic}, where the information is propagated in a widespread manner by diffusing information molecules from a group of randomly moving mobile bio-nanomachines that encounter other bio-nanomachines.

The most interesting aspect in mobile MC is probably the coordinated movement of bio-nanomachines, which allows for the implementation of the functionalities required in practical applications like, for example, cooperative drug delivery or enhanced collective sensing~\cite{soldner2020survey}. In biomolecular monitoring and sensing applications, for instance, sensor bio-nanomachines can detect target chemical molecules and move toward the zone where they are localized~\cite{abdi2017quantization}. In drug delivery applications, bio-nanomachines embedded with drug molecules are injected intravenously into a human or animal body and move toward target sites. The bio-nanomachines communicate and coordinate their motion to search for target sites, aggregate, and release embedded drug molecules~\cite{cao2019diffusive}. In tissue regeneration applications, cell-based bio-nanomachines proliferate, grow, and migrate to facilitate the formation of a tissue structure. For the two mentioned cases, all the receiving bio-nanomachines are able to collectively sense molecules that are released from the target site~\cite{khalid2020modeling}. 


In order to evaluate the performance of MC systems it is therefore needed to develop suitable channel models for scenarios where there are multiple bio-nanomachines that collectively sense the molecules released by a transmitter, i.e. the target site. As an example, the reciprocal effect of two fully-absorbing (FA) receivers was considered in~\cite{two_rx}, where the degradation of the channel is evaluated in terms of Bit Error Rate (BER) and capacity. The same reciprocal effect was exploited in~\cite{regonesi2020relative} to estimate the angle between the direction under which an FA receiver sees the other with respect to segment joining it to the pointwise transmitter. Therefore, a required characteristic of a channel model is that of including the reciprocal effect of the receiving bio-nanomachines. In this paper we focus on an MCvD system defined by a single pointwise transmitter and multiple mobile FA receivers. The main contribution consists in the introduction of an MCvD channel model that takes into account the instantaneous relative position of each receiver with respect to the pointwise transmitter.



\subsection{Related Work and Contributions}

In MCvD a significant number of channel models that take into account the effect of nearby multiple FA receivers have recently appeared in the literature~\cite{guo2016eavesdropper, huang2020channel, kwak2020two, sabu2020detection}. An initial study is done in~\cite{guo2016eavesdropper} for a one-dimensional (1D) model to count discrepancies in the number of molecules absorbed in the case of two FA receivers, with respect to that of a single one. The same 1D environment is then considered in~\cite{huang2020channel} to derive the exact analytical closed-form expression for the fraction of molecules absorbed over time by each receiving nanomachine. A characteristic of the 1D models is that the two receivers and the pointwise transmitter are aligned, with the latter always in the middle. The extension to the three-dimensional (3D) environment with two FA receivers is proposed in~\cite{kwak2020two}, where the fraction of molecules absorbed by each of them is derived. 
A first attempt to infer an analytical formula that takes into consideration the geometry of a 3D scenario with two FA receivers is proposed in~\cite{bao2019channel}. The formula accounts for the effect of the \emph{interferer} \textcolor{black}{(i.e., the unintended receiver)} by introducing a scaling factor on the cumulative number of molecules absorbed by the intended receiver up to a certain time instant $t$, taking into account the geometric scenario. The scaling factor is a function of the geometrical parameters with some adjustment coefficients, specific for each interferer position, obtained by curve fitting of empirical data. A main weakness of the approach is that it assumes that the effect of the interfering receiver can be modelled by a simple reduction in the number of emitted molecules, whereas the actual shape of the channel impulse response can be quite perturbed. Besides, the method can be hardly generalized to an arbitrary number of receivers. 

In this paper we follow a different approach from~\cite{kwak2020two} to characterize the channel impulse response with multiple receivers. We model each absorbing receiver as a source of negative molecules and we apply the superposition principle to obtain the channel response for each transmitter-receiver pair. Our approach leads to an analytical description of the problem as a system of equations that, for the two receivers case, is equivalent to the one proposed in~\cite{kwak2020two}. Compared to this latter, the proposed method has the advantage that it can be easily extended to an arbitrary number of receiving nanomachines, as we will show by means of an example. Besides, it can be generalized to a population of nanomachines of various sizes. The multiple receivers case was already considered in~\cite{sabu2020detection} for a 3D MCvD system, where the centers of the spherical receivers are distributed as a Poisson point process in the medium. However, the analytical expression for the first hitting probability is derived for any of the receivers within time $t$ without referring to any specific position for each receiver. Conversely, as a main contribution of the present paper, we \textcolor{black}{obtain} the channel impulse response of each receiver considering the perturbation introduced by all the interfering nanomachines.
\textcolor{black}{The channel impulse responses of all receivers are constrained by a system of integral equations. We solve the system of equations analytically for} \textcolor{black}{the case of} \textcolor{black}{two receivers. Concerning the scenario with multiple receivers, we switch to numerical integration for which an analytical expression is not available yet. It is worth noting that we use the words \emph{interferer} and \emph{receiver} interchangeably since \textcolor{black}{ the (intended)} receiver} \textcolor{black}{can be considered as an interferer for the other} \textcolor{black}{(unintended)} \textcolor{black}{receiver,} \textcolor{black}{ when} \textcolor{black}{we want to investigate its number of} \textcolor{black}{ absorbed} \textcolor{black}{molecules.}
In addition, we show that, when the interferer is not far from both the transmitter and/or the target receiver, the channel impulse response shape can be quite \textcolor{black}{distorted by its presence}. For instance, the temporal position of the peak in the number of received molecules is shifted, \textit{i.e.}, anticipated or delayed, with respect to the  interferer-free case. The shift is well predicted by the analytical model here proposed.


The paper is organized as follows. In Sec.~\ref{sec:Scenario} we describe the scenario geometry, the motivation, and the analytical results from the literature upon which we build our channel model. In Sec~\ref{section - C-model} we propose a first analytical model, namely the C-model, for the channel impulse response of an MCvD system with one single interferer. \textcolor{black}{Here, the system of integral equations is analytically solved and a closed form expression is provided for the channel impulse response.} In Secs.~\ref{section - S-model} and \ref{section - B-model} we propose two refined models, namely S-model and B-model\textcolor{black}{, and we update the analytical expression found in the preceding section}. In Sec.~\ref{section - multiinterf} we generalize all the models to the scenario of an arbitrary number of interferers, \textcolor{black}{reverting to numerical integration to solve each corresponding system of integral equations.} Finally, in Sec.~\ref{sec:conclusion} we draw our conclusions.
\begin{figure}[!t]
    \centering
    \includegraphics[width=0.55\columnwidth]{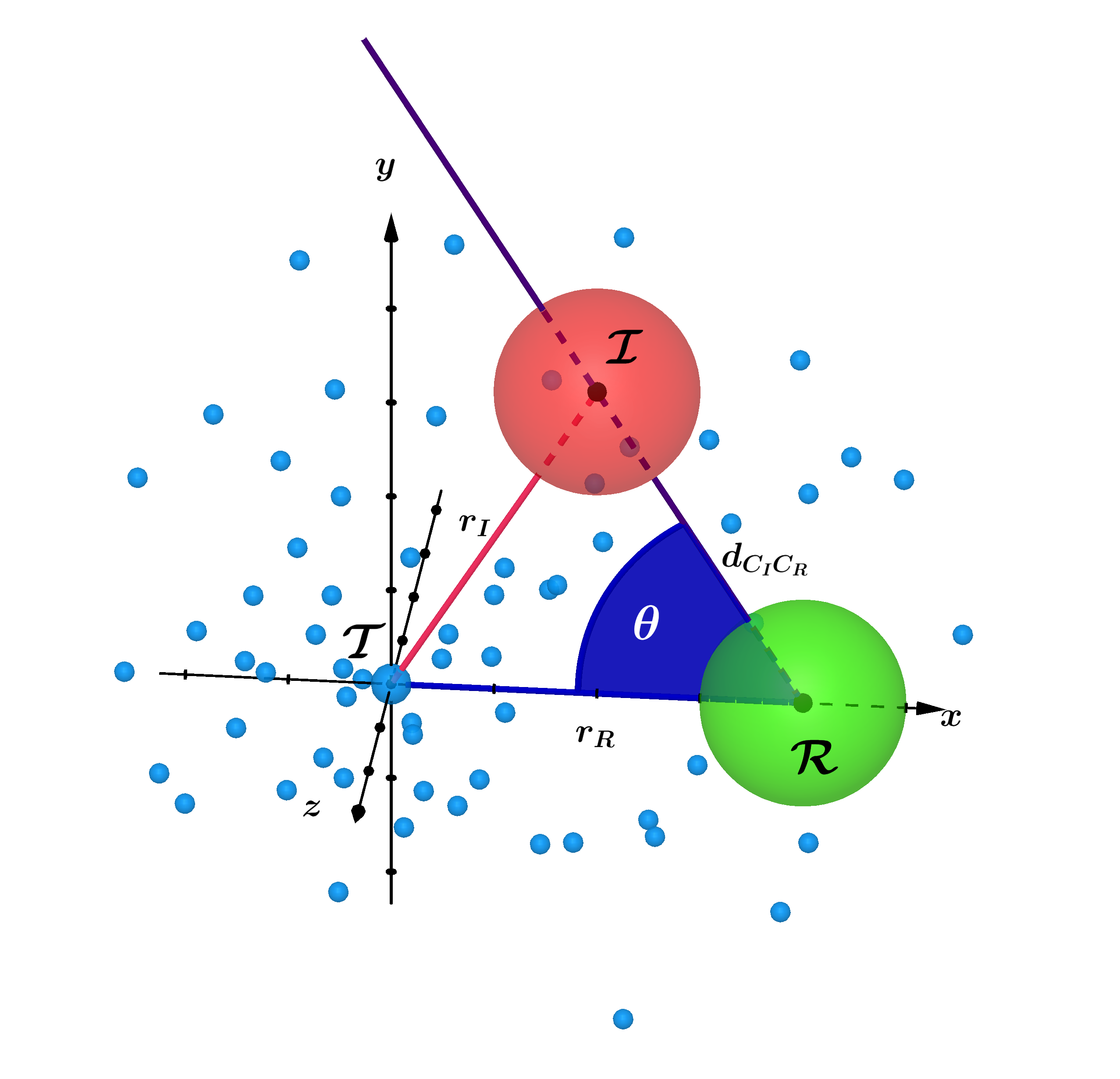}
    \caption{MCvD scenario under study with the pointwise transmitter~($\mathcal{T}$), the receiving~($\mathcal{R}$), and the interfering nanomachine~($\mathcal{I}$).}
    \label{fig:pos_cells}
\end{figure}

\section{System model}
\label{sec:Scenario}
\subsection{Scenario geometry}

A pictorial description of the scenario under investigation is sketched in Fig.~\ref{fig:pos_cells}. A pointwise transmitter $\mathcal{T}$ is placed in the origin and emits impulsively $N_T$ molecules of the same type in the environment at time $t$$\,=\,$$0$. A spherical receiver $\mathcal{R}$,
with radius \mbox{$R$}, is located at distance $r_R$ from $\mathcal{T}$. Possibly
identical interfering receivers are placed arbitrarily in the 3D
space. All receivers are supposed to be of type FA,
\textit{i.e.}, the absorption probability is set
to be \mbox{$p_{abs}=1$}. It follows that, when a molecule $m$ hits the surface of any receiver, it
is absorbed and removed from the environment~\cite{soares2011detection}.

We assume that molecules, supposed to be dimensionless, reach the receiver
through diffusion characterized by a low Reynolds
number~\cite{henry2010introduction}. Thus, the generic molecule
coordinates $(x_m(t),y_m(t),z_m(t))$ at time $t$ can be updated at
the time instant $t+\Delta t$, for small $\Delta t$, according to a random walk law
\begin{align}
    \begin{split}
        x_m(t+\Delta t)&= x_m(t)+\zeta_1 \sqrt{2D\Delta t},\\
        y_m(t+\Delta t)&= y_m(t)+\zeta_2 \sqrt{2D\Delta t},\\
        z_m(t+\Delta t)&= z_m(t)+\zeta_3 \sqrt{2D\Delta t},
    \end{split}
    \label{eq:coord_molec}
\end{align}
where $D$ is the diffusion coefficient, $\Delta t$ is the time step, $\zeta_1$, $\zeta_2$, and, $\zeta_3$ are Gaussian independent random variables, with mean \mbox{$\overline{\zeta}_i=0$} and variance \mbox{$\sigma_{\zeta_{i}}^2=1$} ($i=1,2,3$). 
\renewcommand{\arraystretch}{1.2}

Two scenarios are analyzed in this paper. The first, shown in Fig.~\ref{fig:pos_cells}, is the single-interferer scenario, where $\mathcal{R}$ has its center $C_R$ on the $z$-axis and $\mathcal{I}$ is at distance $d_{C_RC_I}$ from $\mathcal{R}$. Several positions of the interferer are investigated, each one denoted with the angle $\theta$ between the lines joining $\mathcal{T}$ and $\mathcal{R}$ and the one linking the two receivers. The second, which generalizes that of Fig.~\ref{fig:pos_cells}, is the multi-interferer scenario, where all the interferers are placed in a given position at distance $r_{I_k}$ from the origin. On the other side, the relative position of $\mathcal{R}$ with respect to the other cells is changed by varying the angle $\alpha$ between the x-axis and the line conjoining $\mathcal{T}$ and $\mathcal{R}$, being this latter at distance $r_R$ from $\mathcal{T}$.

\textcolor{black}{The case study of this paper can be applied to model the targeted drug delivery where, for example, a swarm of nanomachines containing anticancer drugs swim towards cancer cells and release drugs to kill cancer cells~\cite{zhao2021release}. 
Nanomachines, or modiﬁed bacteria, which absorb and sense a substance released by the target, are the vehicles that transport drug molecules in the delivery site. They} are supposed to be geometrically identical, and their radius is set to $1\,\mu$m. This choice is consistent with real bacteria with coccus cell morphology ({\textit{e.g}. \textit{Staphylococcus} species,
\textit{Streptococcus} species, etc.}), \textit{i.e.}, round-shaped bacteria whose diameter can vary in the range $[0.25,\,1]~\mathrm{\mu m}$~\cite{tortoracl}. The diffusion coefficient $D$ of the propagation medium is $79.4~\mathrm{\mu m^2/s}$, which corresponds to a water diffusive environment in the human body~\cite{kuran2010energy}. Finally, in the following, we consider $N_T=10^4$ released molecules.

\subsection{Dependence of the simulation results on $\Delta t$}

\begin{figure}[!t]
    \centering
    \includegraphics[width=0.5\columnwidth]{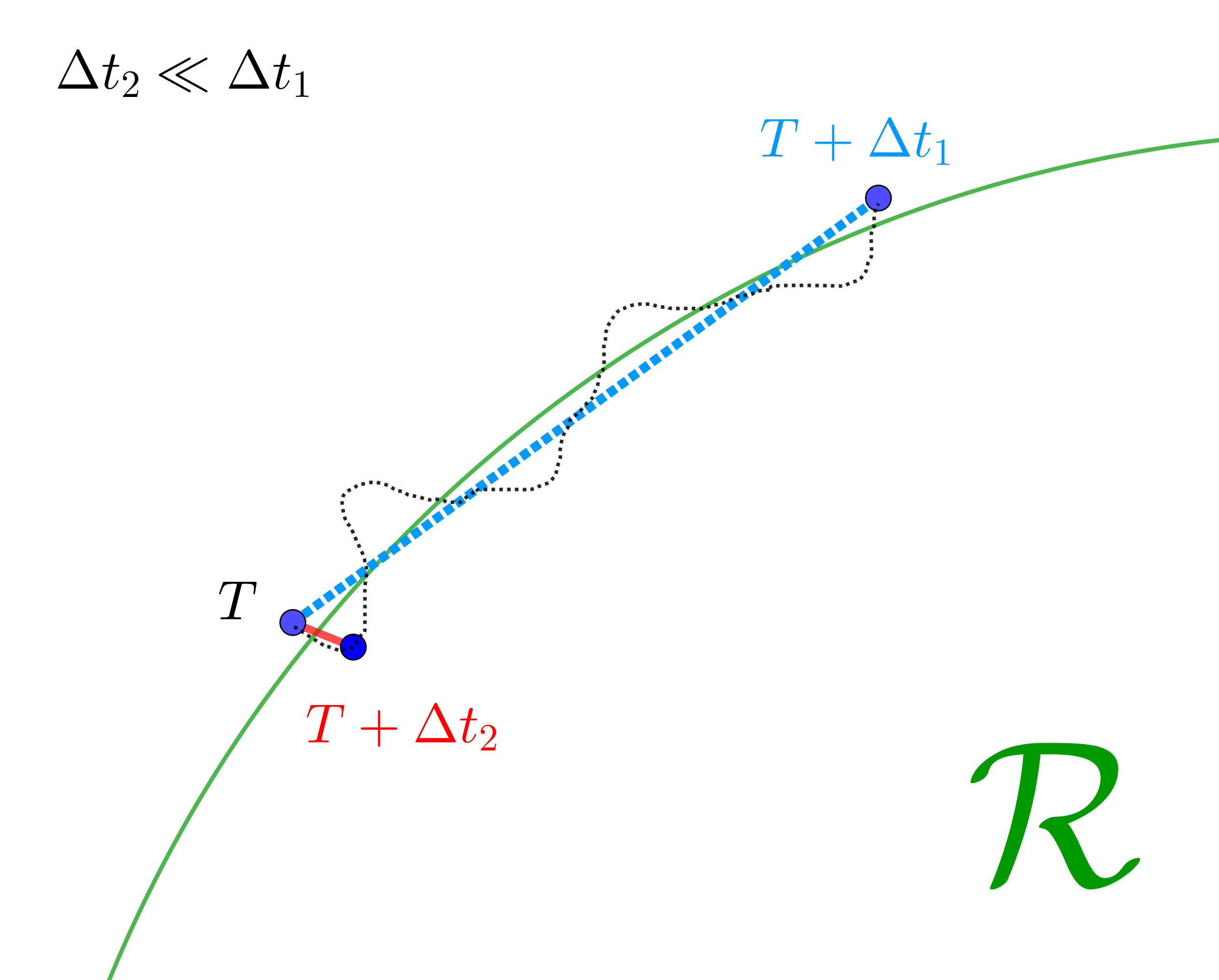}
    \caption{A molecule propagating close to the tangent to the cell surface would not be absorbed if $\Delta t$ were too high.}
    \label{fig:mol_tang}
\end{figure}

A key point in the numerical modelling of an MC system is the choice of the time step $\Delta t$, which influences the granularity of the random walk described by~\eqref{eq:coord_molec}. As a consequence, the continuous-time variation of the position of a molecule is better approximated for low values of $\Delta t$. As shown in Fig.~\ref{fig:mol_tang}, this choice becomes critical when a molecule would cross the cell membrane more than once in a short time interval, for instance, propagating on a trajectory close to the tangent to the surface like the dotted black line: with a large value of $\Delta t$ the update of the molecule position can fail to register its absorption when the time \emph{spent} by it inside the cell is smaller than $\Delta t$.
A simulator, in these cases, would underestimate the correct number of absorbed molecules. 

\begin{figure}[!t]
    \centering
    \includegraphics[width=0.75\columnwidth]{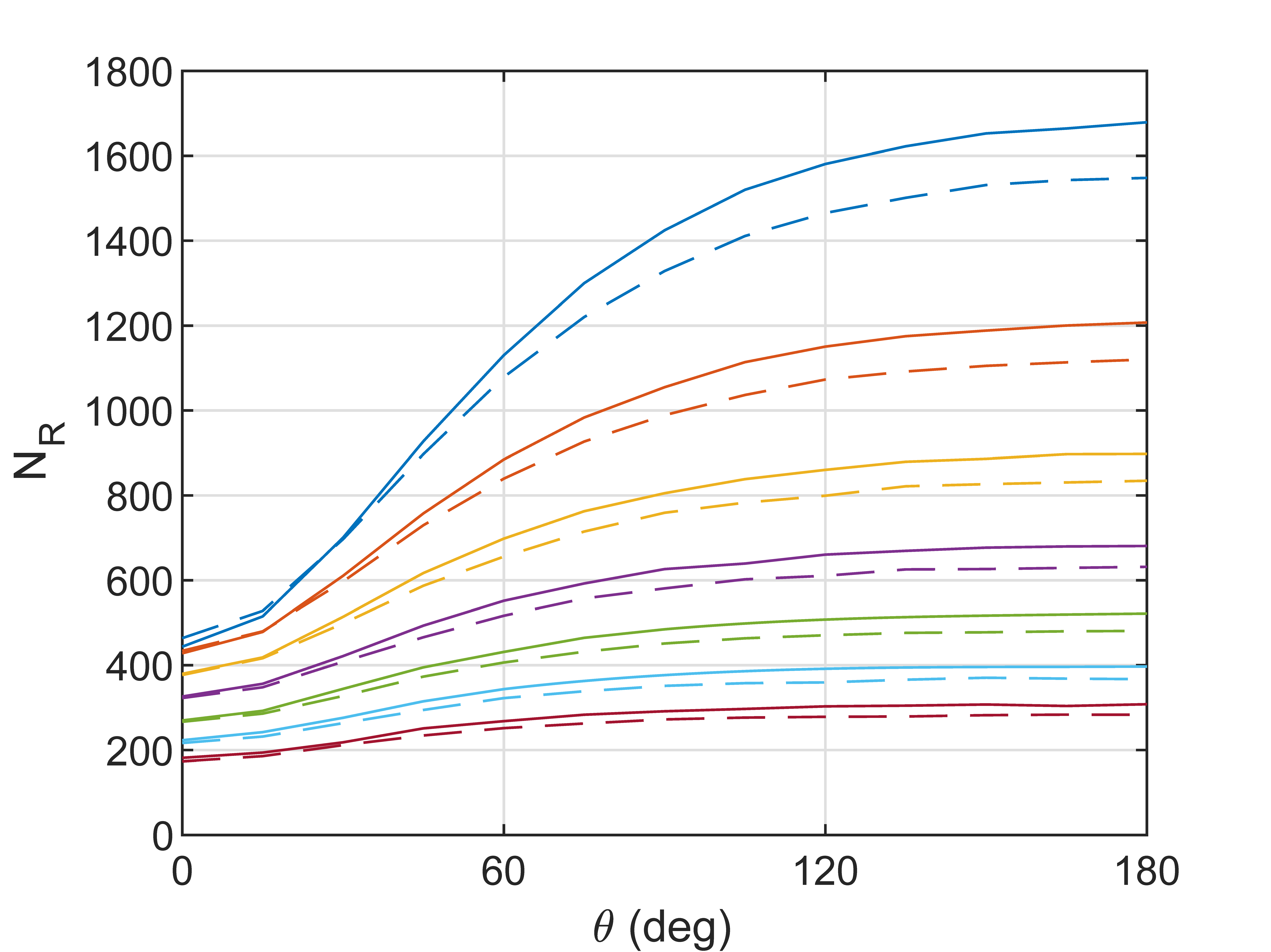}
    \caption{Cumulative number of absorbed molecules for $\Delta t=10^{-6}\,$s (solid lines) and $\Delta t=10^{-4}\,$s (dashed lines). Total simulation time $T=0.5\,$s. From the top to the bottom, $r_R$ increases unitary from $4$ to $10\,\mu$m, for each pair of curves.}
    \label{fig:updated1e_6}
\end{figure}

To show this effect, we consider 
the case where the receiver $\mathcal{R}$ is at distance $r_R$ from $\mathcal{T}$ and an interfering nanomachine $\mathcal{I}$ is at distance $d_{C_I C_R}=r_R/2$ from $\mathcal{R}$. The relative angle $\theta$ under which the receiver sees the interferer with respect to the transmitter assumes values in $\left[0^\mathrm{o}; 180^\mathrm{o}\right]$. For $r_R$ ranging in the interval between $4\,\mu$m and $10\,\mu$m, Fig.~\ref{fig:updated1e_6} shows the number of absorbed molecules $N_R$ after $0.5\,$s with $\Delta t = 10^{-4}\,$s and $\Delta t = 10^{-6}\,$s. For low values of $\theta$ and small distances $r_R$, the number of molecules absorbed by $\mathcal{R}$ decreases when $\Delta t = 10^{-6}\,$s: $\mathcal{I}$ absorbs more molecules, and, being interposed between $\mathcal{T}$ and $\mathcal{R}$ and very close to $\mathcal{R}$, its blocking effect is more relevant.
When $r_R$ increases, so does $d_{C_IC_R}$, and the effect of reducing $\Delta t$ is an increase of the molecules absorbed at $\mathcal{R}$. The same happens when $\theta$ increases. At $r_R$$\,=\,$$4\,\mu$m the molecules absorbed by the target receiver are decreased of $4.5\,\%$ at $\theta$$\,=\,$$0^\mathrm{o}$ and they are increased of $7.5\,\%$ at $180^\mathrm{o}$. When $r_R\textbf{}=~10\,\mu$m, $\mathcal{R}$ absorbs $4.5\,\%$ more molecules at $\theta = 0^\mathrm{o}$ and $8\,\%$ at $180^\mathrm{o}$.
On the other side, by decreasing the value of $\Delta t$, the number of samples increases, leading to a higher computation time. So a trade-off between accuracy and time has to be found.
The estimation of the absorbed molecules when $\Delta t$ varies is not straightforward in presence of two or more receivers due to the interplay among them. For this reason, in even more complex scenarios, an analytical model for the channel impulse response would be of utmost importance. 

\subsection{Single transmitter/single receiver channel impulse response} \label{subsec:P2P channel imp. resp.}

If $\mathcal{T}$ were alone in the 3D space, or the receiving
nanomachines were transparent, the emitted molecules would propagate, with perfect spherical symmetry, according to the microscopic theory of diffusion which is ruled by the concentration gradient \cite{akkaya2014effect,jamali2019channel}.
The Fick's second law in a 3D environment binds the time
derivative of the flux to the Laplacian of the molecule concentration $p\left(r,t\right)$ as
\begin{equation}
\frac{\partial p \left( r,t \right)}{\partial t} = D \nabla^2p
\left( r,t \right). \label{eq - 2nd Fick}
\end{equation}

A transparent receiving cell would sense in each point the
molecule concentration $p \left( r,t \right)$ which solves
\eqref{eq - 2nd Fick} with the boundary conditions
\begin{align}
    \begin{split}
        \lim_{t \rightarrow 0} p \left( r,t \right) &= Q \delta \left( r \right),
        \\
        \lim_{r \rightarrow \infty} p \left( r,t \right) &= 0,
    \end{split}
    \label{eq:coord_interf}
\end{align}
\textit{i.e.},
\begin{equation}
p \left( r,t \right) = \frac{Q}{\sqrt{4 \pi D t^3}}
e^{-\frac{r^2}{4Dt}}. \label{eq - free space f}
\end{equation}

On the contrary, the presence of any, non-transparent receiving
cell perturbs the spherical symmetry of the problem by capturing
molecules from the environment. The concentration must still obey
\eqref{eq - 2nd Fick}, but it takes a more general form where $p
\left( \mathbf{x},t \right)$ depends on the 3D coordinate point
vector $\mathbf{x}$ rather than solely on the distance $r$ from
the origin. The perturbation of the spherical symmetry is due to
the boundary conditions that account for the absorbing behavior of
the receiving cell. 
In \cite{yilmaz2014three} the problem with a single FA receiver is smartly solved by transforming it into an
equivalent problem where the spherical symmetry is preserved. 
The equivalence does not apply in general to the concentration $p
\left( r,t \right)$ but it holds as far as the \emph{first hit
probability} is concerned. This allows to conclude that the
analytical expression of the \emph{hitting rate} of the molecules
onto the receiving cell surface, namely $f \left( r,t \right)$,
also holds for the case of a concentrated transmitter.
The
expression, which can be considered the \emph{impulse response} of
an MCvD channel with a single FA receiver $\mathcal{R}$, of radius
$R$ and centered at distance $r_R$ from $\mathcal{T}$, reads
\begin{equation}
f \left( r_R,t \right) = \frac{R \left(r_R-R\right)}{r_R\sqrt{4 \pi D
t^3}} e^{-\frac{\left(r_R-R \right)^2}{4Dt}}, 
\label{eq:imp_res}
\end{equation}
and the \emph{absorption rate}, \textit{i.e.}, the number of molecules absorbed by the cell per unit time, is
\begin{equation}
n_R \left( t \right) = N_T f \left( r_R,t \right) 
\label{eq:n_R(t)}
\end{equation}
when the transmitter $\mathcal{T}$ emits $N_T$ molecules impulsively. The number of absorbed molecules is obtained by integrating \eqref{eq:n_R(t)} up to time instant $t$ as
\begin{equation}
    N_R(t)=\int_{0}^t n_R \left( u \right) \,du.\label{eq:integ}
\end{equation}

\subsection{Perturbation introduced by an interfering nanomachine}
\begin{figure}[!t]
    \centering
    \includegraphics[width=0.7\columnwidth]{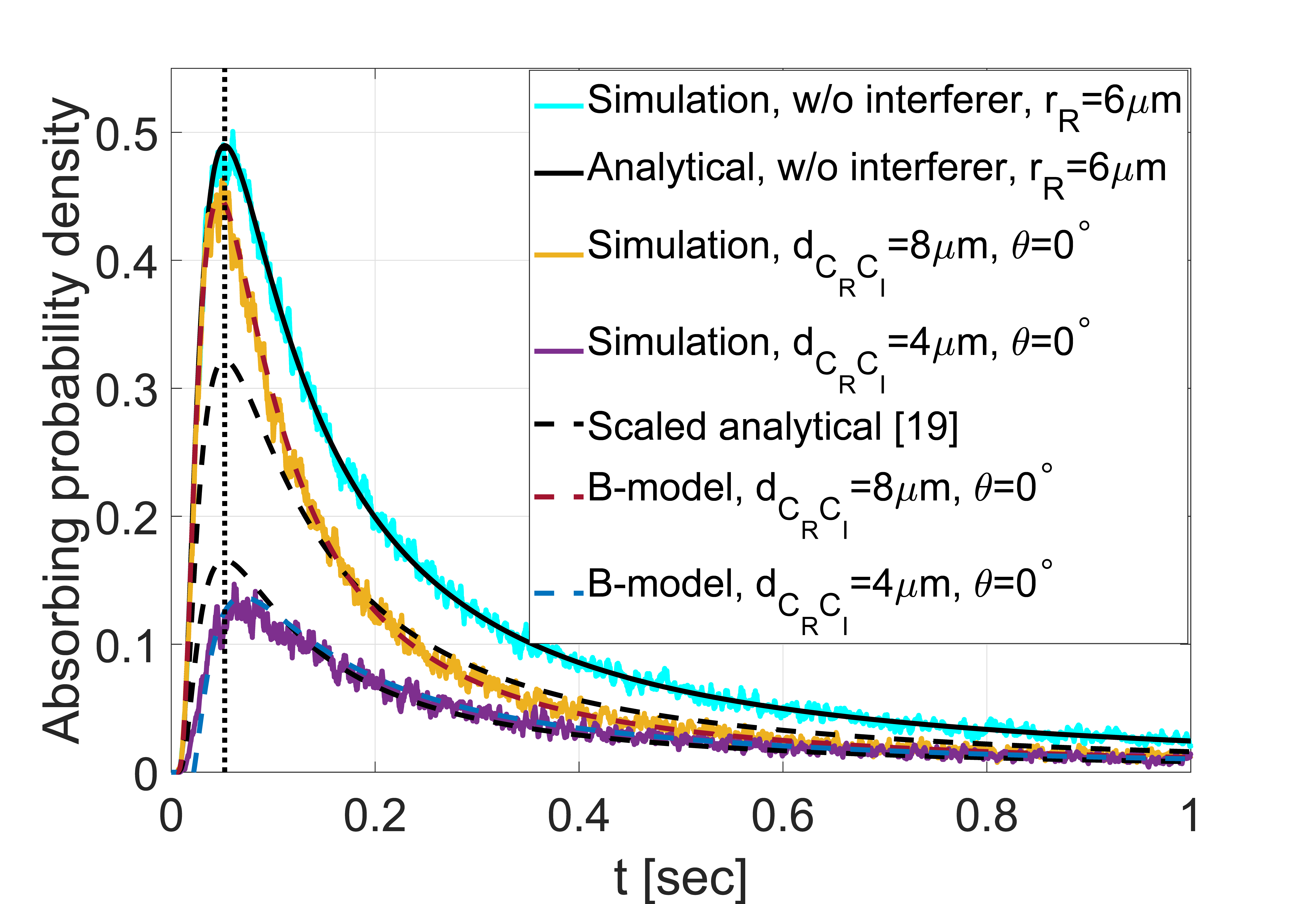}
    \caption{Comparison between simulated and predicted analytical impulse responses obtained for an interferer placed at the two different sides of the pointwise transmitter  on the line that joins it with center of the spherical nanoreceiver. The temporal shift of the peak is well predicted by the analytical B-model given in Sec.~\ref{section - B-model}.}
    \label{fig:lessaveraged}
\end{figure}

The presence of an FA interferer that subtracts molecules to the target receiver perturbs the channel impulse response \eqref{eq:imp_res}. Unlike assumed in \cite{bao2019channel}, the perturbed response is not just a version of \eqref{eq:imp_res} reduced in amplitude, since it is also modified in shape.
For instance, we have observed in our particle-based simulation that the assumption that the position of the peak time is not affected by the presence of the interferer can be considered as a good approximation when the interferer is far from both the transmitter and/or the target receiver. In such a case, as first observed in~\cite{bao2019channel}, the effect of the interferer consists in a simple scaling of the impulse response between the pointwise transmitter and the spherical receiver. Actually, in our particle-based simulations we observed that, when such a condition is not true, there is a shift in the position of the peak time. Figure~\ref{fig:lessaveraged} shows the absorbing probability density versus time for the case where the intereferer is located along the line that connects the transmitter and the receiver. Two cases are shown in the figure: i) the interferer is between the transmitter and the receiver; ii) the interferer is on the side that is opposite to the target receiver with respect to the transmitter. To show the shift in the position of the peak time we followed the same procedure of~\cite{bao2019channel} by comparing the results obtained by particle-based simulation with the analytical curves.
\begin{figure}[!t]
    \centering
    \includegraphics[width=0.7\columnwidth]{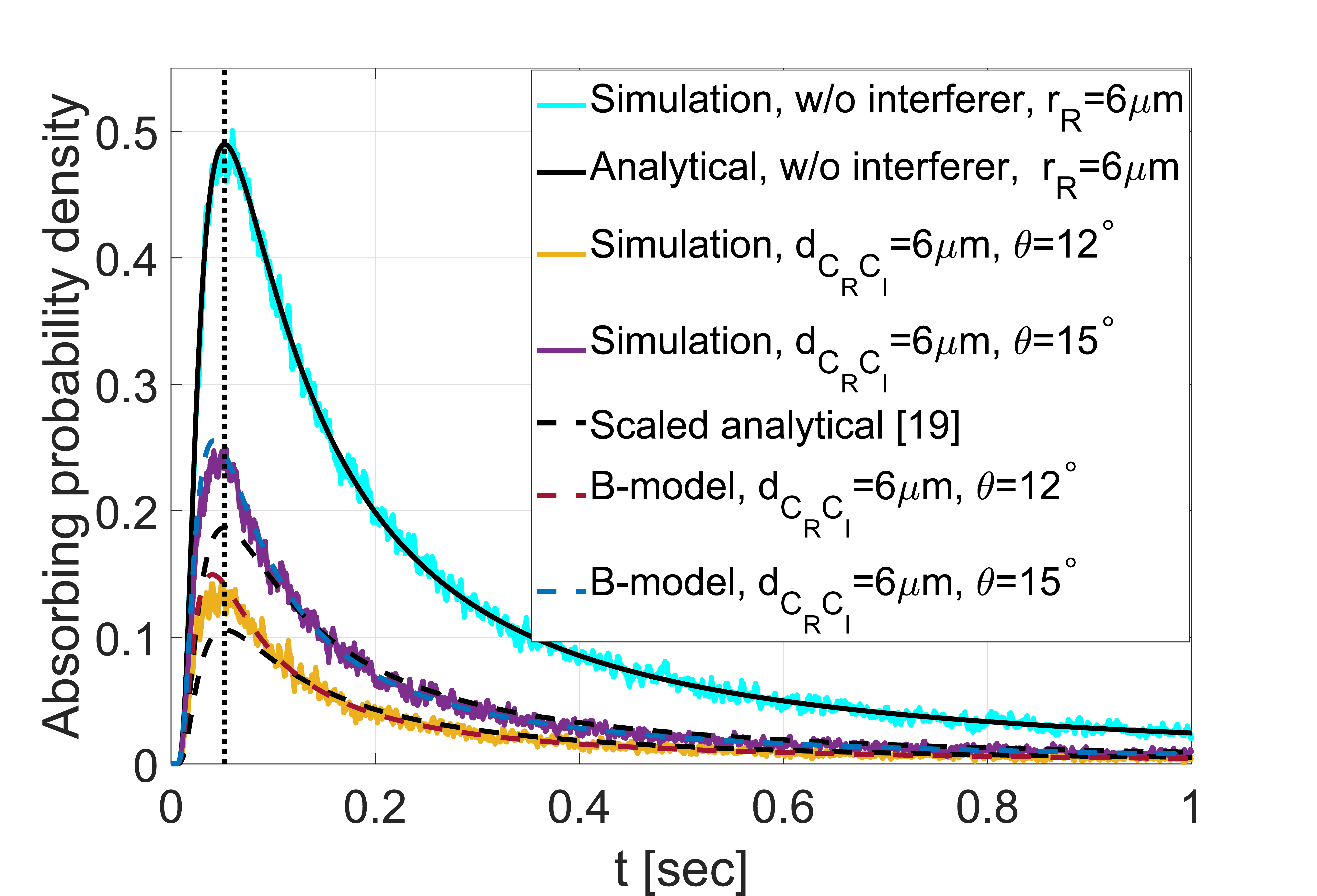}
    \caption{Comparison between simulated and predicted analytical impulse responses obtained in the case of an interferer in two different positions close to the pointwise transmitter at a fixed distance from the receiver. The temporal shift of the peak is well predicted by the analytical B-model given in Sec.~\ref{section - B-model}.}
    \label{fig:lessaveragedangle}
\end{figure}

Figure~\ref{fig:lessaveraged} shows that, when the interferer lies between the transmitter and the receiver, the peak time is shifted ahead with respect to that predicted by the theory. This is due to the presence of a blocking effect that is introduced by the interferer on the receiver, which causes a delay with respect to the situation where it is not present, \textcolor{black}{thus leading to a postponed peak compared to that predicted by the theory in case of a single transmitter with a single receiver. In contrast,} when the interferer is behind and close to the transmitter, it starts absorbing immediately the molecules. \textcolor{black}{Its effect arrives at the receiver with delay and results in} a faster decrease in the number of molecules in the environment, \textcolor{black}{thus leading to an anticipation of the position of the peak time compared to that predicted by the theory for the case of a single receiving nanomachine.} In turn, $\mathcal{R}$ observes a decrease in the number of absorbed molecules before to that predicted by the theory. As a reference, we also report the comparison in the case of absence of the interferer, where a perfect match is observed between simulation and analytical curve. 

To further emphasize the effect of the interferer on the position of the peak, in Fig.~\ref{fig:lessaveragedangle} we report the numerical results for the case where $\mathcal{I}$ is at the same distance as the one between the receiver and the transmitter with $\theta$$\,=\,$$12^\circ, 15^\circ$, \textit{i.e.}, the interferer is close to the pointwise transmitter (similar considerations would hold if it were close to the receiver). In this case the anticipation of the peak time is more evident compared to the case considered in Fig.~\ref{fig:lessaveraged}, where it is behind the transmitter. 

Both in Figs.~\ref{fig:lessaveraged} and \ref{fig:lessaveragedangle} we also anticipate the absorbing probability densities predicted by one of the models proposed in this paper, namely the B-model, described in detail in Sec.~\ref{section - B-model}. As it can be observed, the agreement with simulation is excellent, and the peak time shifts of the curves are accurately predicted.

\section{Center interferer model}
\label{section - C-model}

Consider the case of a single FA interferer $\mathcal{I}$ as in Fig.~\ref{fig:pos_cells}, for the time being. Without loss of generality we assume that it is centered in $C_I$, at distance $r_I$ from $\mathcal{T}$ and at distance $d_{C_IC_R}$ from $\mathcal{R}$. If $\mathcal{T}$ emits $N_T$ molecules impulsively at time $t=0$, both receiving cells would sense an absorption rate \eqref{eq:n_R(t)}, each with its own distance $r_R$ or $r_I$, if they were alone. To compute $\mathcal{R}$'s absorption rate
$n_R\left(t\right)$ in presence of the interferer $\mathcal{I}$ we
invoke the superposition principle. The hitting rate \eqref{eq:imp_res} must be combined with
an expression that solves the second Fick's law \eqref{eq - 2nd
Fick} and satisfies the additional boundary condition at
$\mathcal{I}$, which absorbs molecules with an (unknown) absorption rate $n_I\left(t\right)$. The effect of this absorption can be accounted for as a negative
source, since molecules are subtracted from the environment. The
effect of this source signal, which we assume concentrated in the
center $C_I$ of $\mathcal{I}$ for the moment, propagates by diffusion
and perturbs $n_R\left(t\right)$ according, again, to the channel impulse
response \eqref{eq:imp_res}.
Obviously, the
distance $r$ in this case is the distance
$d_{C_IC_R}$ between $\mathcal{R}$ and $\mathcal{I}$.

Finally, by the symmetry of the problem, the same reasoning can be
applied exchanging the roles of the absorption rates
$n_R\left(t\right)$ and  $n_I\left(t\right)$, \textit{i.e.},
\begin{equation}
    \begin{cases}
        n_R\left(t\right) = N_T f \left( r_R,t \right) - n_I\left(t\right) \star f \left( d_{C_IC_R},t \right)
        \\
        n_I\left(t\right) = N_T f \left( r_I,t \right) - n_R\left(t\right) \star f \left( d_{C_IC_R},t \right)
    \end{cases},
    \label{eq - C-model equations}
\end{equation}
\textcolor{black}{where $\star$ denotes the convolution. The system of integral equations \eqref{eq - C-model equations} can be analytically solved in time domain, since the channel impulse response $f\left( r,t \right)$ is causal.}
We refer to this solution as the C-model for the channel impulse response, since the absorbing, negative source is concentrated in the center $C$ of each absorbing cell.
\textcolor{black}{We derive the number of observed molecules at receiver $\mathcal{R}$ from integration of~\eqref{eq - C-model equations}, as follows (see Appendix~\ref{Ap:proof1G})
 \begin{align}
     \resizebox{0.8\hsize}{!}{$\begin{aligned}
    N_{R}\left( t\right)& = \frac{N_{T}R}{r_{R}} \sum_{n=0}^{\infty} \left(\frac{R}{d_{C_{I}C_{R}}}\right)^{2n} \mathrm{erfc}\left ( \frac{r_{R}-R  + 2n\left( d_{C_{I}C_{R}}-R \right)}{2\sqrt{Dt}} \right ) \\
    &\hspace{0.4cm}- \frac{N_{T}R}{r_{I}} \sum_{n=0}^{\infty} \left(\frac{R}{d_{C_{I}C_{R}}}\right)^{2n+1} \mathrm{erfc}\left ( \frac{r_{I}-R  + \left ( 2n+1 \right ) \left( d_{C_{I}C_{R}}-R \right)}{2\sqrt{Dt}} \right ),
    \end{aligned}$} \label{eq:SITO_N1_det}
 \end{align}
where 
\begin{equation}
    \mathrm{erfc}\left ( z \right ) = 1 - \frac{2}{\sqrt{\pi}}\int_{0}^{z} e^{-\tau ^2}d\tau
\end{equation}
is the complementary error function.} \textcolor{black}{Note that} \textcolor{black}{\eqref{eq:SITO_N1_det} converges to~\eqref{eq:integ} by moving the interferer far from the receiver~$\mathcal{R}$, \textit{i.e.,} observation of the receiver will be the same as if there was no interferer (see Appendix~\ref{Ap:proof2}).
The channel impulse response between the transmitter and the receiver in the presence of an interferer can be derived from~\eqref{eq:SITO_N1_det}
 \begin{align}
     \resizebox{0.9\hsize}{!}{$\begin{aligned}
    f\left(r_{R},r_{I},d_{C_{R}C_{I}}, t\right)& = \frac{R}{r_{R}} \sum_{n=0}^{\infty} \left(\frac{R}{d_{C_{I}C_{R}}}\right)^{2n} \bigg ( \frac{r_{R}-R  + 2n\left( d_{C_{I}C_{R}}-R \right)}{2\sqrt{\pi Dt^3}} e^{-\frac{\left(r_{R}-R  + 2n\left( d_{C_{I}C_{R}}-R \right)\right)^2}{4Dt}}\bigg )\\
    &\hspace{0.4cm}- \frac{R}{r_{I}} \sum_{n=0}^{\infty} \left(\frac{R}{d_{C_{I}C_{R}}}\right)^{2n+1} \bigg ( \frac{r_{I}-R  + \left ( 2n+1 \right ) \left( d_{C_{I}C_{R}}-R \right)}{2\sqrt{\pi Dt^3}} e^{-\frac{ \left(r_{I}-R  + \left ( 2n+1 \right ) \left( d_{C_{I}C_{R}}-R \right)\right)^2}{4Dt}} \bigg)
    \end{aligned}$}\hspace{0.2cm}, \label{eq:SITO_CIR}
 \end{align}
and
\begin{equation}
    N_R(t)=\int_{0}^t N_{T}f\left(r_{R},r_{I},d_{C_{R}C_{I}}, u\right)  \,du .
\end{equation}
}
\begin{figure*}[!t]
    \centering
    \includegraphics[width=\columnwidth]{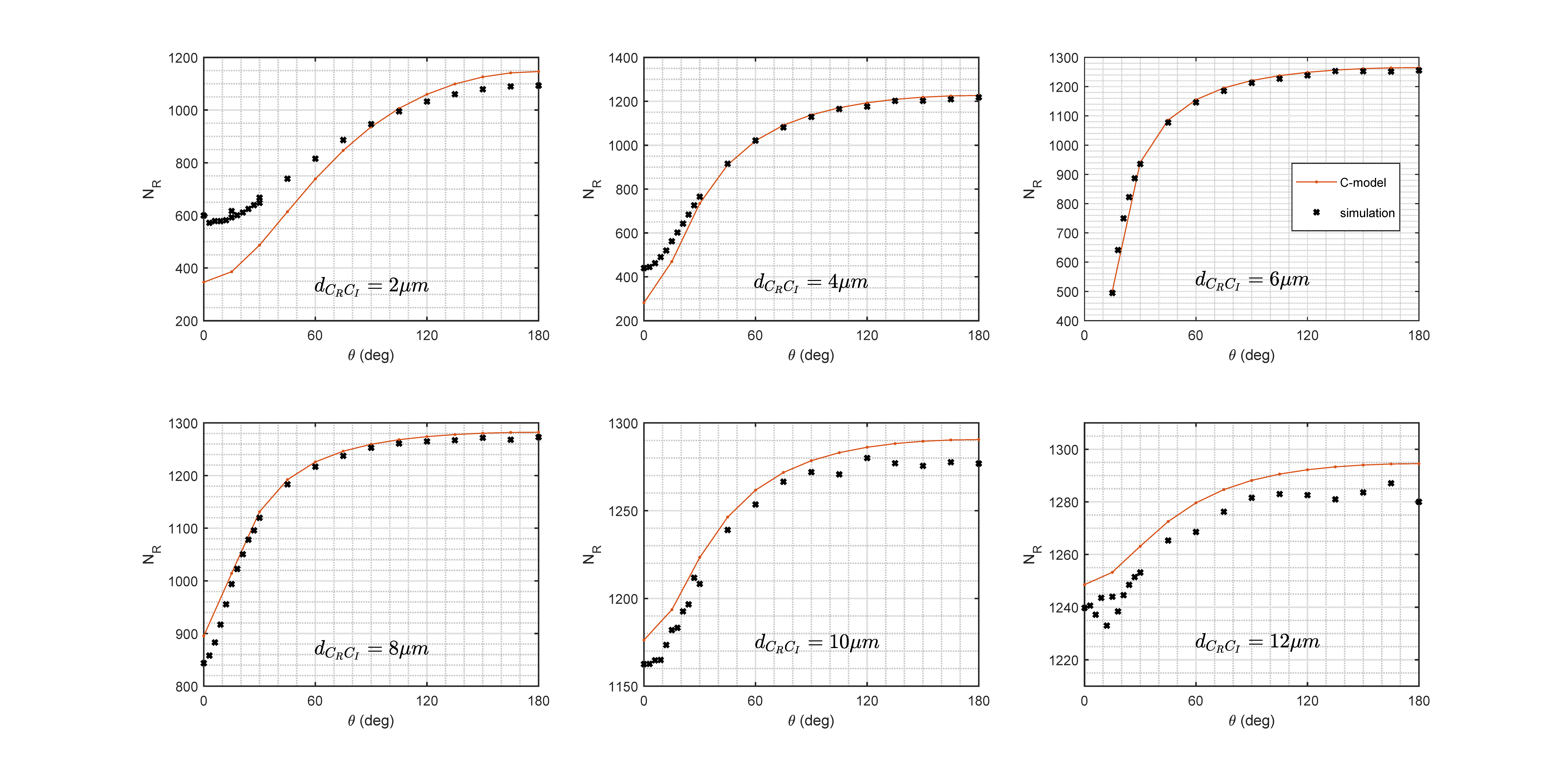}
    \caption{Cumulative number of molecules $N_R(T)$ absorbed by $\mathcal{R}$ after $T=2$ s, in the scenario
     of Fig. \ref{fig:pos_cells} with $r_R=6\,\mu\mathrm{m}$, for various
     positions of $\mathcal{I}$ identified by $d_{C_RC_I}$ and $\theta$. C-model prediction versus simulation $\left( \Delta t = 10^{-6}\,\mathrm{s}\right)$.}
    \label{fig - model C}
\end{figure*}

Intuitively, the solution can be seen in this way. By the
memoryless nature of Brownian motion, the probability that a given
molecule $m$, placed at point $P$ at time $t$, hits the receiver
$\mathcal{R}$, within the next $T-t$ seconds, is independent of
how $m$ reached the point $(P,t)$ and depends solely on its
distance from $\mathcal{R}$ and on the observation time $T-t$.
Thus, the absorption of any molecule by $\mathcal{I}$ subtracts
it from the amount of possible molecules that will be able to hit
$\mathcal{R}$, starting from the surface of $\mathcal{I}$ at time
$t$. Taken by itself, $\mathcal{I}$ acts as a negative source of
molecules whose effect on the hitting rate on $\mathcal{R}$ via
diffusion must be filtered through the channel impulse response
$f\left( r,t \right)$.
It is worth observing that the system of equations in~\eqref{eq - C-model equations} is consistent with~\cite[eq. (29)]{kwak2020two}, where an equivalent system, obtained through a probabilistic approach, is described in its integral form.
\begin{figure}[t!]
    \centering
    \includegraphics[width=0.8\columnwidth]{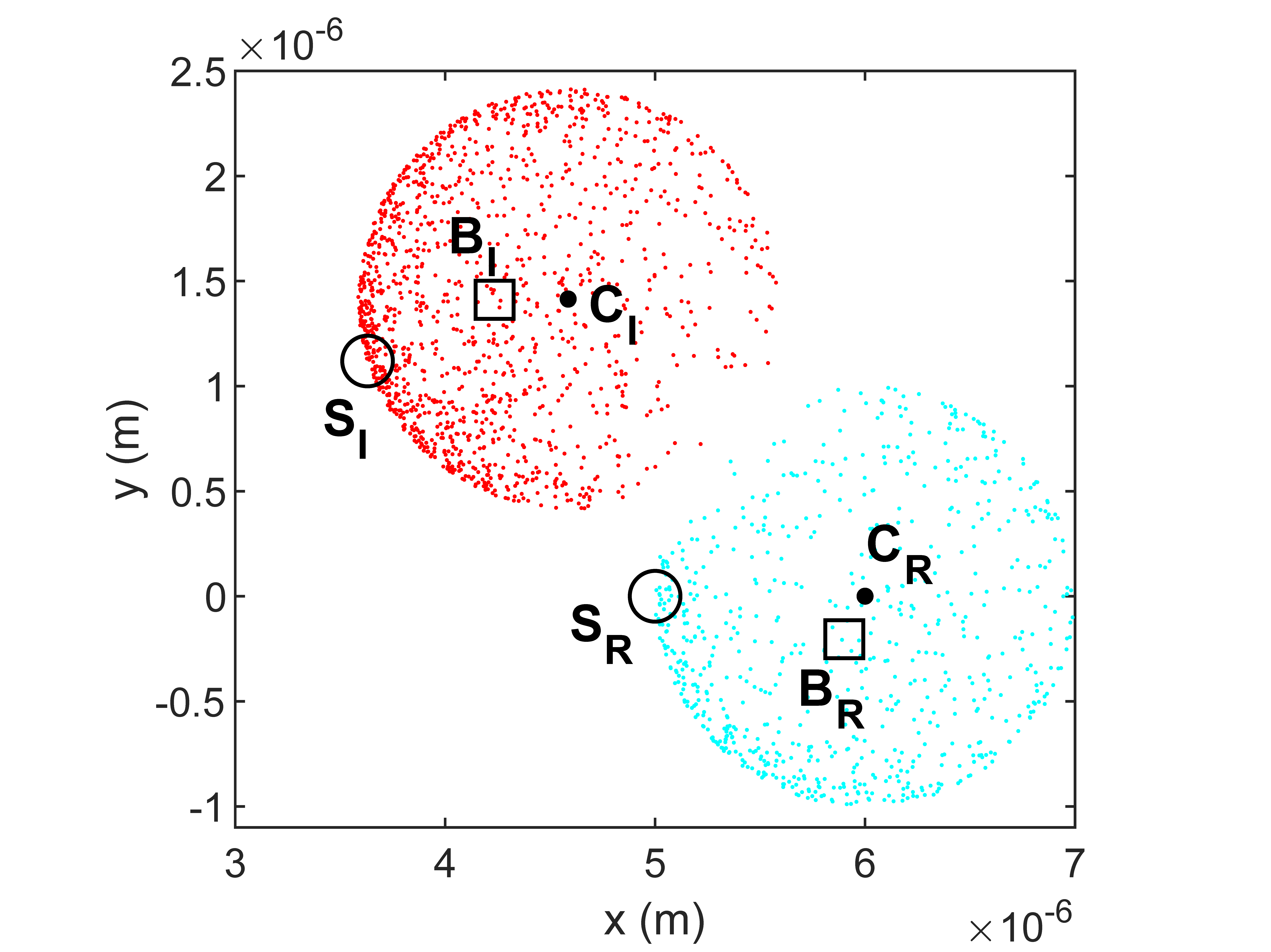}
    \caption{2-D representation of the numerical assessment of the absorbed molecules barycenters (black squares), for both $\mathcal{I}$ and $\mathcal{R}$. The black dots represent the cell centers, the black circles represent the cell boundary points closer to the transmitter. The fine colored points are the positions of absorbed molecules in each cell.}
    \label{fig - barycenter assessment on example simulation}
\end{figure}

In Fig.~\ref{fig - model C} we compare the C-model prediction with
simulated data in terms of number of absorbed molecules $N_R$ after $2\,$s for various scenarios, where $\mathcal{I}$ is placed at various distances $d_{C_I C_R}$ from $\mathcal{R}$ and various angles $\theta$. The agreement is good when the distance between $\mathcal{I}$ and $\mathcal{T}$ is large (large $d_{C_IC_R}$ or large $\theta$). On the contrary, the model predictions fail when $\theta$ is small, in particular for small $d_{C_IC_R}$ as well\footnote{Note that the $y$-axis range of the subplots in Fig.~\ref{fig - model C} for large $d_{C_IC_R}$ is much smaller than for small $d_{C_IC_R}$, since $N_R$ is less sensitive to $\theta$.}. 


The assumption of the C-model that $\mathcal{I}$ acts as a source
of negative molecules, concentrated in the center $C$ of the cell,
is certainly an approximation. By the intuitive interpretation
proposed above, the correct position of the source should be
assumed distributed on $\mathcal{I}$'s surface where the molecules are
absorbed. This observation explains why the analytical model
perfectly matches the simulation results when the distances are
large with respect to the cells radius, whereas its predictions are less
accurate when the cells are close each other. Furthermore, it
introduces the investigation described in the next section.

%
%
%

\section{Surface interferer model}\label{section - S-model}

Consider the scenario in Fig.~\ref{fig - barycenter assessment on example simulation} where the distances between $\mathcal{T}$, $\mathcal{R}$, and
$\mathcal{I}$ are comparable with the cells radius. According to
the C-model, the molecules captured by $\mathcal{I}$ and removed
from the environment in any point of $\mathcal{I}$ surface, are
taken into account by modelling a source of negative molecules
concentrated in the center $C_I$ of $\mathcal{I}$. This leads to a
distortion since the molecules are actually destroyed in places
that are possibly closer to (or sometimes even further apart from)
the surface of $\mathcal{R}$.

As a first guess the portion of the surface of $\mathcal{I}$ that
has the highest probability of being hit, is its side that is
facing $\mathcal{T}$. As a first correction trial (S-model), we
can thus assume that the negative, absorbing source, instead of
being concentrated in the center $C_I$ of $\mathcal{I}$, is
actually placed in the point $S_I$ of the surface boundary of
$\mathcal{I}$ which is the closest to $\mathcal{T}$, namely the
\emph{S-point} of $\mathcal{I}$, which is marked by a circle in
Fig.~\ref{fig - barycenter assessment on example simulation}. The only thing that needs to be
changed in \eqref{eq - C-model equations} is that the distance
$d_{C_IC_R}$ must be replaced by $d_{S_IC_R}$.
The same correction is applied when considering the effect of
$\mathcal{R}$ on $\mathcal{I}$. Note that, in general, the
distance $d_{S_IC_R}$ considered in this case is
different from $d_{C_IS_R}$.
\begin{equation}
    \begin{cases}
        n_R\left(t\right) = N_Tf \left( r_R,t \right) - n_I\left(t\right) \star f \left( d_{C_RS_I},t \right)
        \\
        n_I\left(t\right) = N_Tf \left( r_I,t \right) - n_R\left(t\right) \star f \left( d_{C_IS_R},t \right)
    \end{cases} 
    \label{eq - S-model equations}
\end{equation}

\textcolor{black}{
We derive the number of observed molecules at receiver $\mathcal{R}$ from integration of~\eqref{eq - S-model equations}, as follows (see Appendix~\ref{Ap:proof1G})
 \begin{align}
     \resizebox{.9\hsize}{!}{$\begin{aligned}
    N_R(t) &= \frac{N_{T}R}{r_R}\sum_{n=0}^{\infty}\frac{R^{2n}}{(d_{C_{R}S_{I}}d_{C_{I}S_{R}})^{n}} \mathrm{erfc}\left(\frac{(r_R-R)+n(d_{C_{R}S_{I}}+d_{C_{I}S_{R}}-2R)}{2\sqrt{Dt}}\right) \\
    &\hspace{0.4cm}- \frac{N_{T}R^{2}}{d_{C_{R}S_{I}}r_I}\sum_{n=0}^{\infty}\frac{R^{2n}}{(d_{C_{R}S_{I}}d_{C_{I}S_{R}})^{n}} \mathrm{erfc}\left(\frac{(d_{C_{R}S_{I}}+r_I-2R)+n(d_{C_{R}S_{I}}+d_{C_{I}S_{R}}-2R}{2\sqrt{Dt}}\right)
     \end{aligned}$} .\label{eq:SITO_N1_detG}
 \end{align}
}
\begin{figure*}[t!]
    \centering
    \includegraphics[width=\columnwidth]{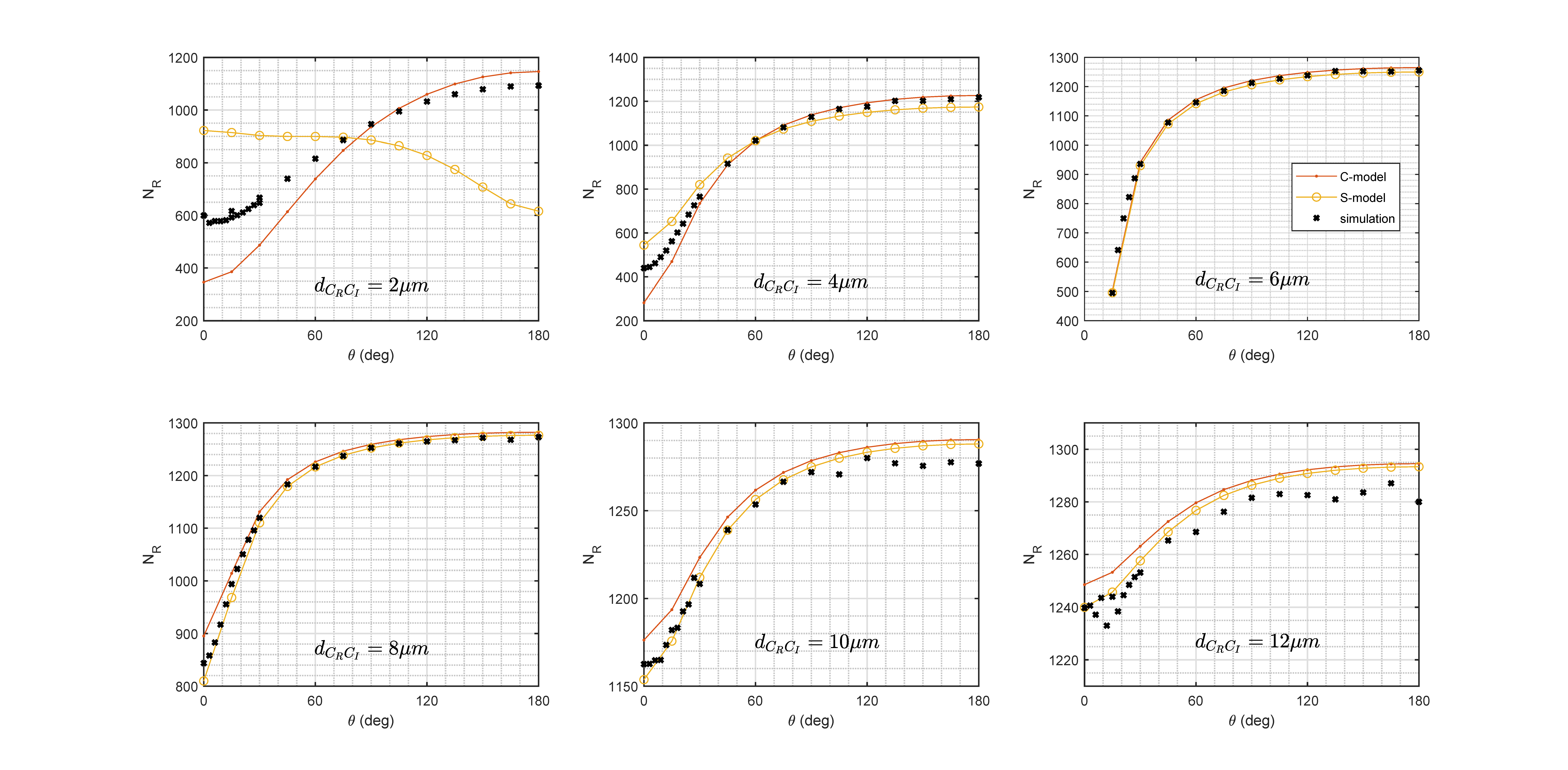}
    \caption{Cumulative number of molecules $N_R(T)$ absorbed by $\mathcal{R}$ after $T=2$ s, in the scenario
     of Fig. \ref{fig:pos_cells} with $r_R=6\,\mu\mathrm{m}$, for various
     positions of $\mathcal{I}$ identified by $d_{C_RC_I}$ and $\theta$. C-model and S-model predictions versus simulation $\left( \Delta t = 10^{-6}\,\mathrm{s}\right)$.}
    \label{fig - C-model and S-model}
\end{figure*}

In Fig.~\ref{fig - C-model and S-model} we compare the S-model and
C-model predictions with simulation. When the two model
predictions agree, \textit{i.e.}, when the distances are large
with respect to the cells radius, the models both match the simulation
results. In the other cases, the simulation results lie in between
the two predictions, which means that when the C-model
overestimates the effect of the interferer, the S-model
underestimates it and vice-versa. This observation introduces the
investigation described in the next section.

\section{Barycenter interferer model}
\label{section - B-model}

The observation that concludes Sec.~\ref{section - S-model}
raises the idea that the real behavior of the system is somehow in
between the C-model and the S-model. In this section we study a more
appropriate position for the concentrated source of molecules to
replace the dot and circle points of Fig.~\ref{fig - barycenter assessment on example simulation} which lead to the C-model and S-model, respectively.
Such a position should model the fact that molecules are absorbed
by the membrane of the cell with a distribution that is not
uniform over the membrane surface: we want to localize the
\emph{absorption barycenter} (or simply \emph{barycenter}, hereinafter), \textit{i.e.}, the spatial average of the membrane points hit by the
molecules.

\subsection{Absorbed molecules barycenter analysis}\label{subsec:Absorbed molecules barycenter analysis}

The barycenter of the absorbed molecules is expected to be
time-variant. In general it is tightly coupled with the
transmitted waveform and it depends on the time observation $T$, on
the cell position and on the presence of possible interferers.
Sticking to the impulsive transmission hypothesis of
$\mathcal{T}$, it is reasonably close to the S-point of the cell
boundary, for small $T$ and it progressively shifts towards the center of the cell, as $T$ increases
and more molecules surround the cell and are absorbed.
%

Furthermore, a deeper look to the first cases of Fig.~\ref{fig - C-model and S-model}, \textit{i.e.}, when the distance $d_{C_IC_R}$ is low, shows
that the cumulative number of absorbed molecules $N_R$ is placed
in a midway between the predictions of the two models for low
receiver-to-interferer angles $\theta$. In such cases, the barycenter is expected to be in the middle between the center and the S-point of the cell.
As long as $\theta$ increases, and the interfering cell moves
behind (with respect to the transmitter) the receiver cell, the
numeric evaluation fairly agrees with the C-model. 

The barycenter has been assessed from the
numerical analysis. An example case, with $d_{C_IC_R}=2\,\mu$m and $\theta = 45^{\circ}$, is shown in Fig.~\ref{fig - barycenter assessment on example simulation}, where the small colored dots inside each cell are the projection on the $xy$ plane of the membrane points
hit by molecules within the first $T=2\,$s of simulation.
First, it can be noted that, in proximity of the cells contact
point, the absorbed molecules are rare. Second, the computed
barycenters, marked with squares, do not necessarily belong to
the $C-S$ segment, instead they somehow repel each other.
\begin{figure}[t!]
    \centering
    \includegraphics[width=0.75\columnwidth]{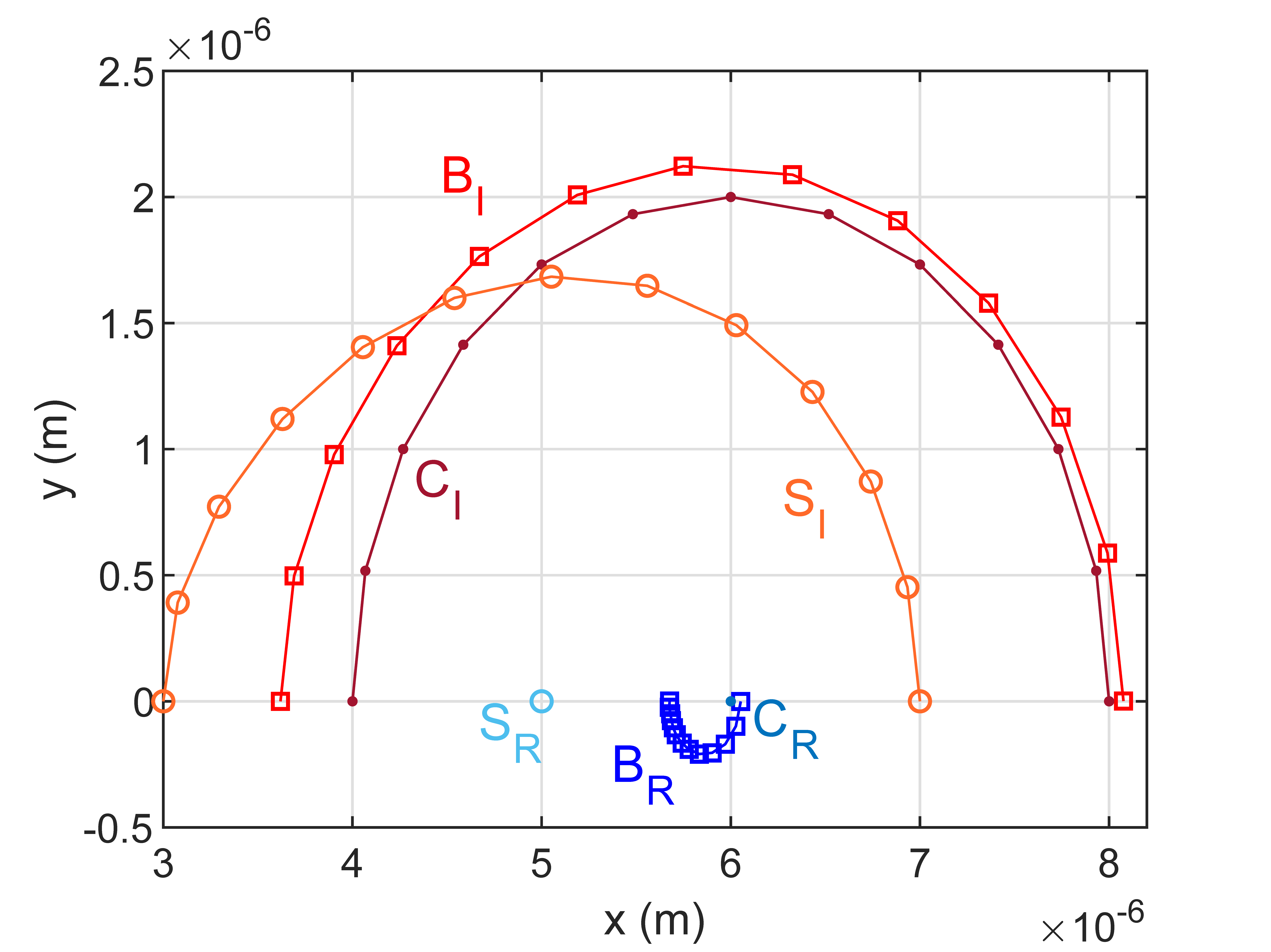}
    \caption{2-D representation of the \textcolor{black}{final position} of the absorbed molecules barycenters (squares), the cell centers (dots), and the cell closest points to the transmitter (circles), for both $\mathcal{I}$ and $\mathcal{R}$.}
    \label{fig - barycenter modelling}
\end{figure}

The plot in Fig.~\ref{fig - barycenter modelling} shows the
evolution of the barycenters in the $xy$ plane when the interferer
$\mathcal{I}$ revolves at distances $d_{C_IC_R}\,=\,2,4,6\,\mu$m around the
receiver $\mathcal{R}$ centered in $C_R=(6,0,0)\,\mu$m. The blue squares are the positions of the $\mathcal{R}$ barycenter $B_{R}$. 
Since $\mathcal{R}$ is static and fixed in space, the shift of the
estimated absorbed molecules barycenter $B_{R}$ position may be
ascribed to a mutual effect with $\mathcal{I}$. The two different
light blue points $S_{R}$ (circle) and $C_{R}$ (dot) represent the receiver 
S-point and center, respectively. The red squares are the barycenters $B_{I}$ of $\mathcal{I}$. Also note
from this plot the mutual repulsion effect, since the $B_{I}$
curve slightly lays beyond the interferer center (dark red) for
mid-to-high angles $\theta$.


\subsection{Absorbed molecules barycenter modelling}

\begin{figure}[t!]
    \centering
    \includegraphics[width=0.75\columnwidth]{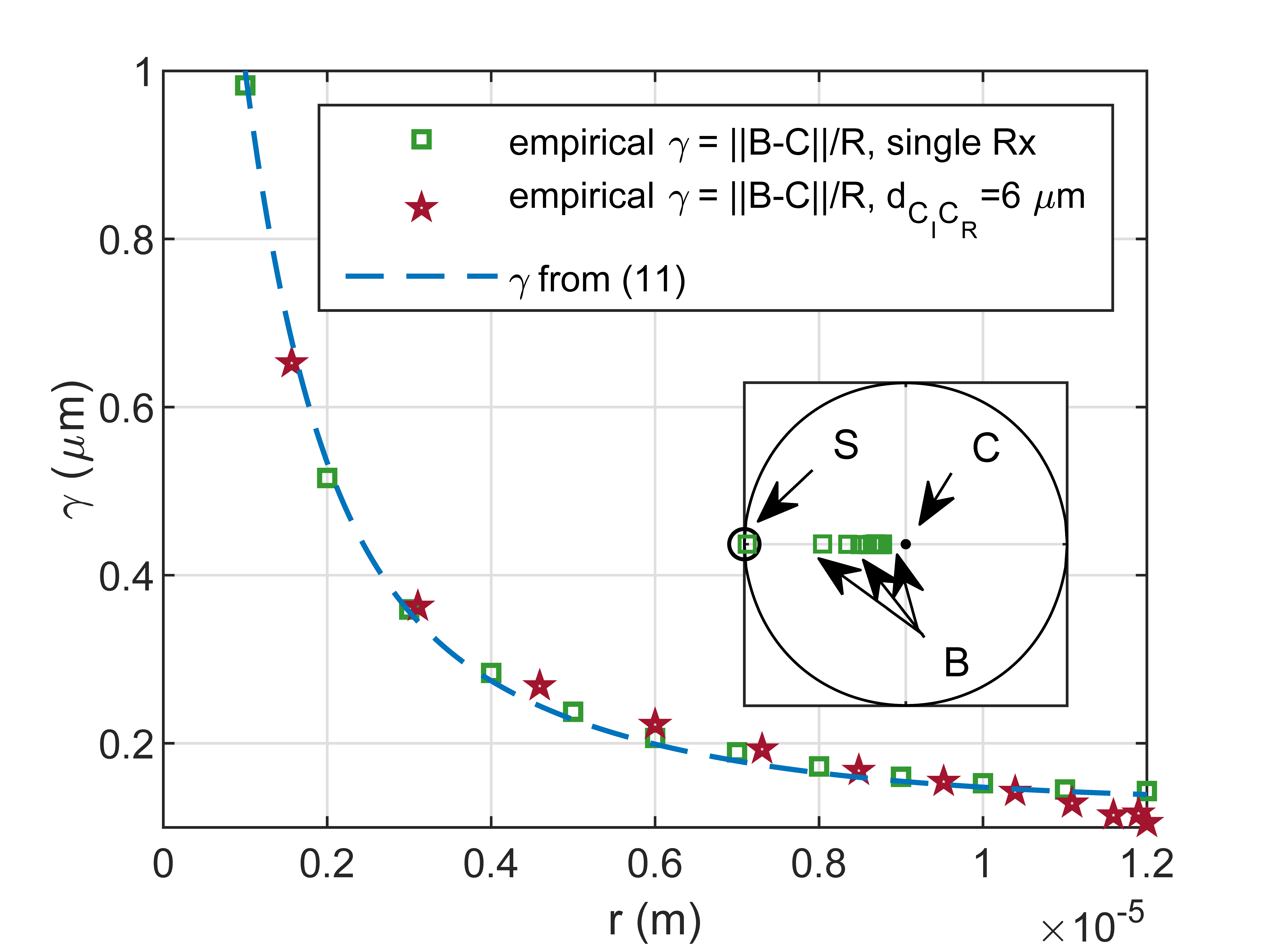}
    \caption{Plot of the distance (normalized by $R$) between the barycenter $B$ and the
    center $C$ of a cell placed at distance $r$ from the origin ($\mathcal{T}$).
    The exponential function $\gamma \left(r\right)$ given in \eqref{eq: gamma3}
    assumed to predict the position of $B$ is also shown.}
    \label{fig - gamma}
\end{figure}

\begin{figure*}[t!] 
    \centering
  \subfloat[\label{1a}]{%
       \includegraphics[width=0.5\columnwidth]{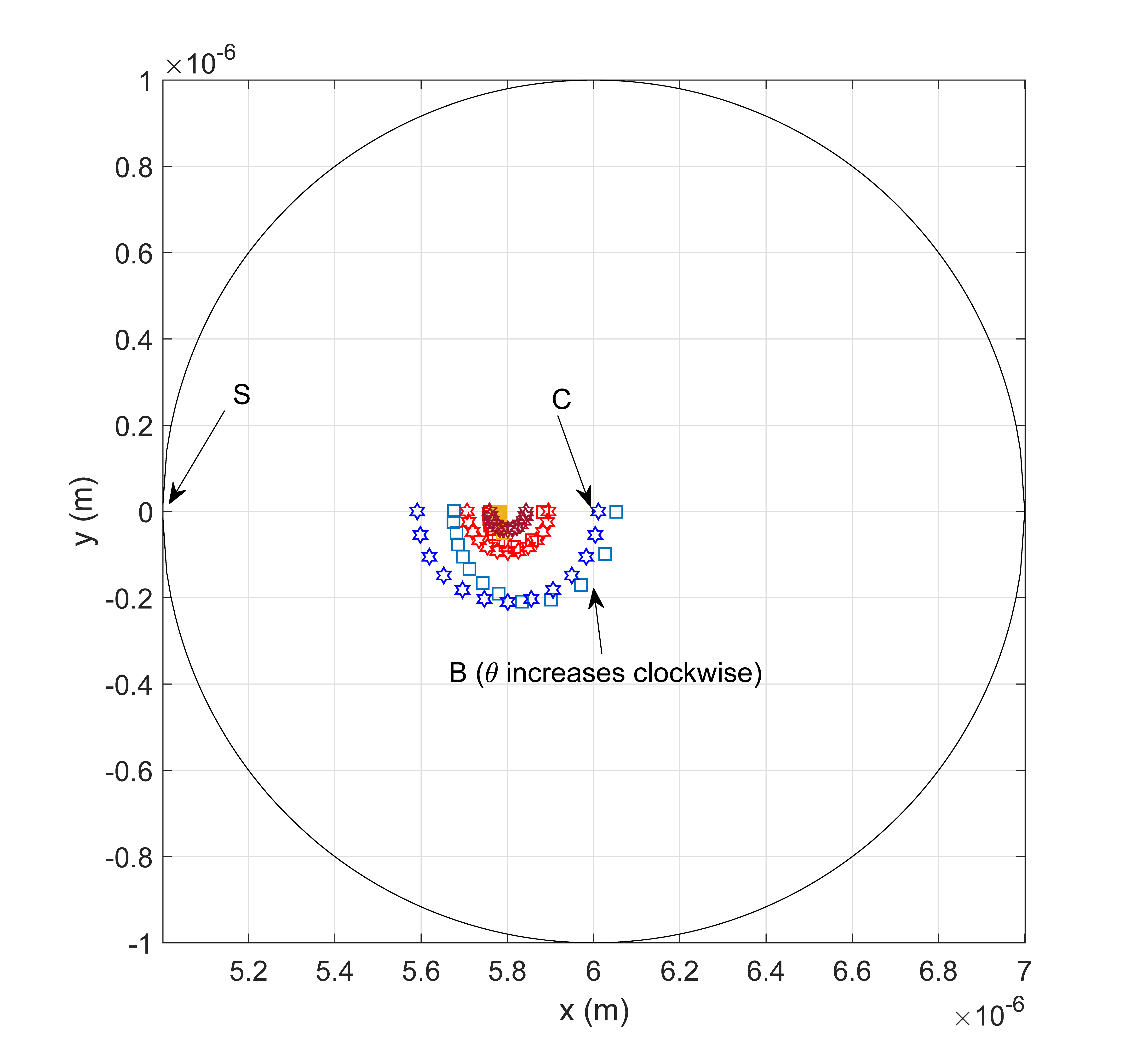}}
  \subfloat[\label{1b}]{%
        \includegraphics[width=0.5\linewidth]{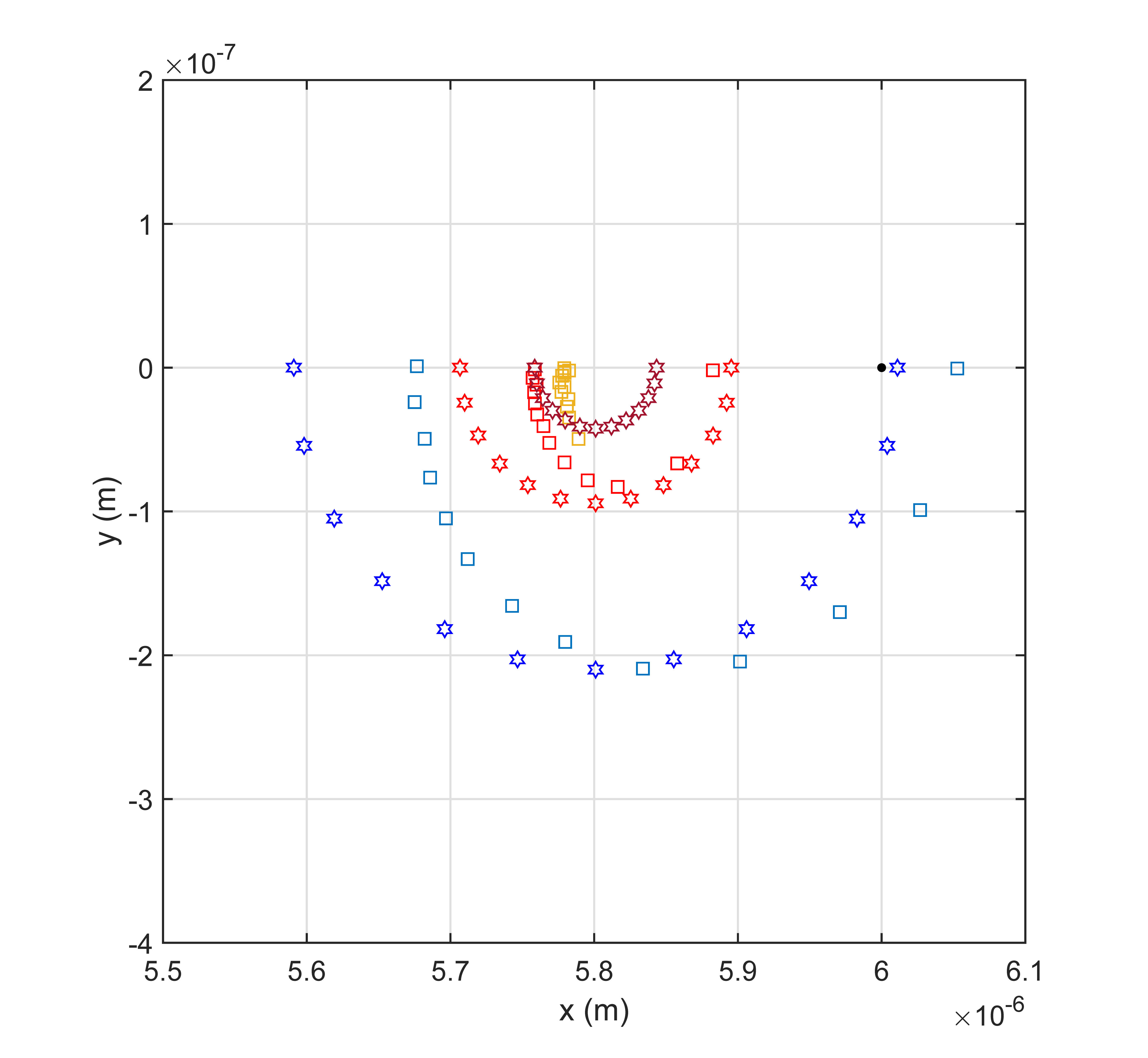}}
  \caption{(a) Position of the barycenter $B_R$ (squares) \textcolor{black}{inside the spherical receiver} $\mathcal{R}$ measured when the interferer $\mathcal{I}$ revolves around $\mathcal{R}$ at distances $d_{C_I C_R}=2$ (blue), $4$
    (red) and $6\,\mu$m (brown). The positions predicted by the model are also shown as hexagons. (b) Magnified position of the barycenter in the proximity of the center of the spherical receiver.}\label{fig - b_R}
\end{figure*}


Based on the preceding empirical observations, we model the
barycenter position $B$ combining two contributions. The first
one, namely $B_0$, does not depend on the presence of other
interfering cells, and differs from the cell center $C$ because of
the non-uniform concentration of molecules in the space
surrounding the cell. As such, it presumably depends on the
diffusion coefficient $D$ and on the observation time $T$ as
stated but, for the time being with fixed $D$ and $T$, we want to
stress its dependence on the distance $r$ of the cell center from
the transmitter (due to the spherical symmetry of the molecules
concentration).

In order to neglect the mutual repulsion effect we measured through simulation the barycenter position of a single receiving cell centered at distance $r$ from $\mathcal{T}$, with $r$ ranging from $1$ to $12\,\mu$m. In Fig. \ref{fig - gamma} we show how the position of the barycenter $B$ shifts progressively from $S$ towards the cell center $C$ as $r$ increases and we plot by circles the distance between $B$ and $C$,
normalized by the cell radius $R$, versus $r$. The figure clearly shows that such a distance
decreases exponentially with increasing $r$. Identifying this
normalized distance as $\gamma$ we can predict the barycenter
position as\footnote{When used inside algebraic expressions, 3D points as $S,C,B_0$ are meant as 3D vectors of real coordinates.},
\begin{equation} 
  B_{0}=\gamma S + (1-\gamma) C.
\label{eq:b_0}
\end{equation}

%

In Fig. \ref{fig - gamma} we show an analytical law for
$\gamma$, obtained through simulated data interpolation through a double exponential function, \textit{i.e.},
\begin{equation}
  \gamma = 0.13 + 0.51 \exp \left( -\frac{r}{0.8 R} \right) + 0.36 \exp \left( -\frac{r}{3 R} \right).
\label{eq: gamma3}
\end{equation}
Also shown marked with stars are the measured normalized distances $||B_I-C_I||/R$ of $\mathcal{I}$ for the case with $d_{C_I C_R}=6~\mu m$ (the largest one, in order to neglect the mutual repulsion effect) for all $\theta$. The comparison with the other curves confirms that the assumption that $\gamma$ is independent of the presence of other interferers is well fit.

The second contribution to the localization of the barycenter aims at modelling the repulsion effect due to the reciprocal effect of shadowing between the two cells. This effect, as already noted, is more relevant
when the two cells are close each other and vanishes with
increasing distance between the cells. We model this effect
through a \emph{displacement} vector $\boldsymbol{\delta}$ of the barycenters in the
direction joining the two cell centers, \textit{i.e.}, for $B_R$, for
instance,
\begin{equation}
  \frac{\boldsymbol{\delta}\left(C_R,C_I\right)}{||\boldsymbol{\delta}\left(C_R,C_I\right)||} =
  \frac{C_R-C_I}{||C_R-C_I||}=-\frac{\boldsymbol{\delta}\left(C_I,C_R\right)}{||\boldsymbol{\delta}\left(C_I,C_R\right)||}.
\label{eq: delta}
\end{equation}

\begin{figure}[t!]
    \centering
    \includegraphics[width=0.75\columnwidth]{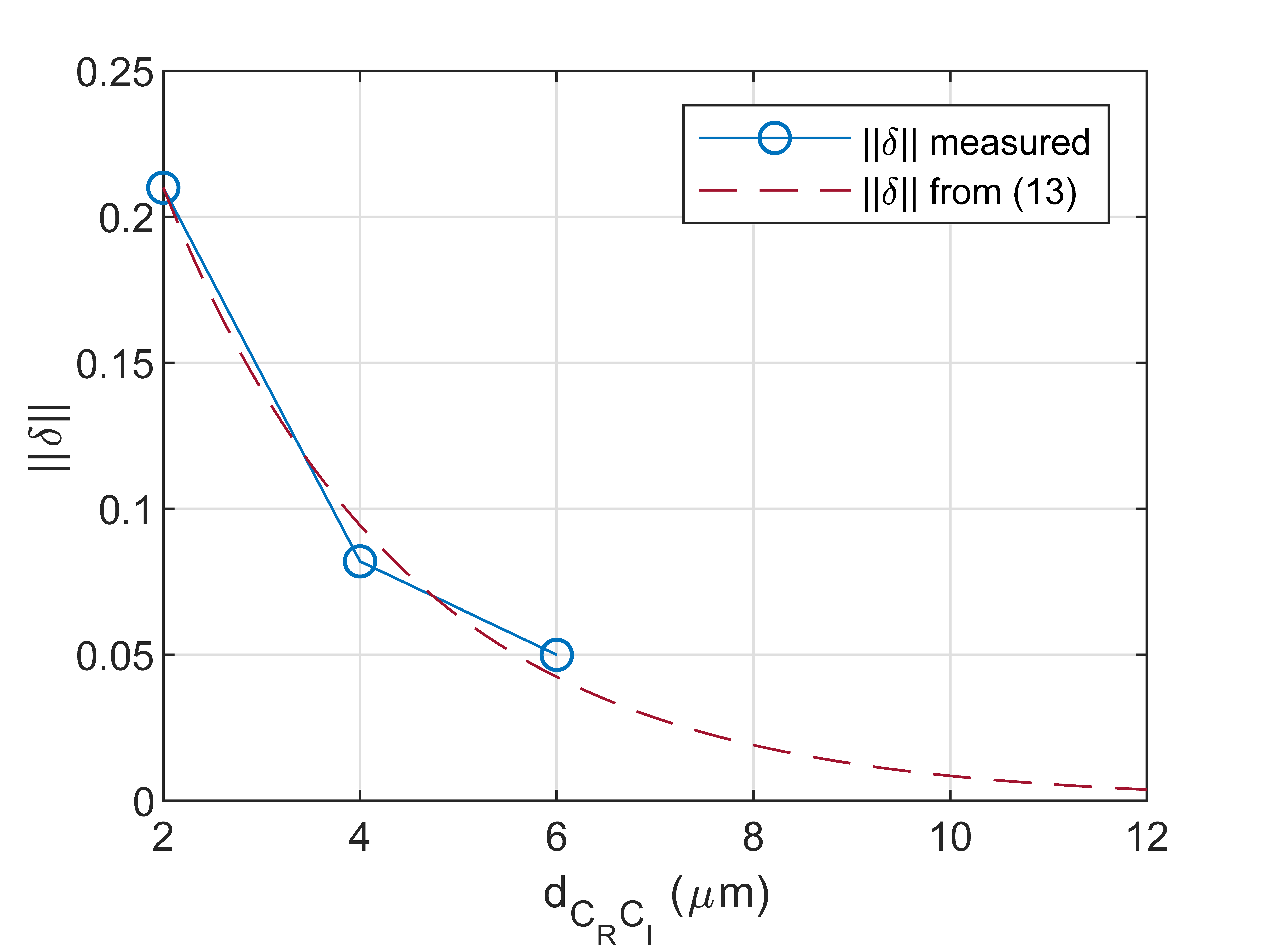}
    \caption{Plot of the norm of the measured displacement $\boldsymbol{\delta} \left(C_R,C_I \right)$ of the
    barycenter $B_R$ from $B_0$ given by \eqref{eq:b_0}, as a function of the distance
    $d_{C_I C_R}$. The exponential function given in \eqref{eq: deltaExponentialLaw}
    assumed to model $||\boldsymbol{\delta}||$ is also shown.}
    \label{fig - delta}
\end{figure}

\begin{figure}[t!]
    \centering
    \includegraphics[width=0.75\columnwidth]{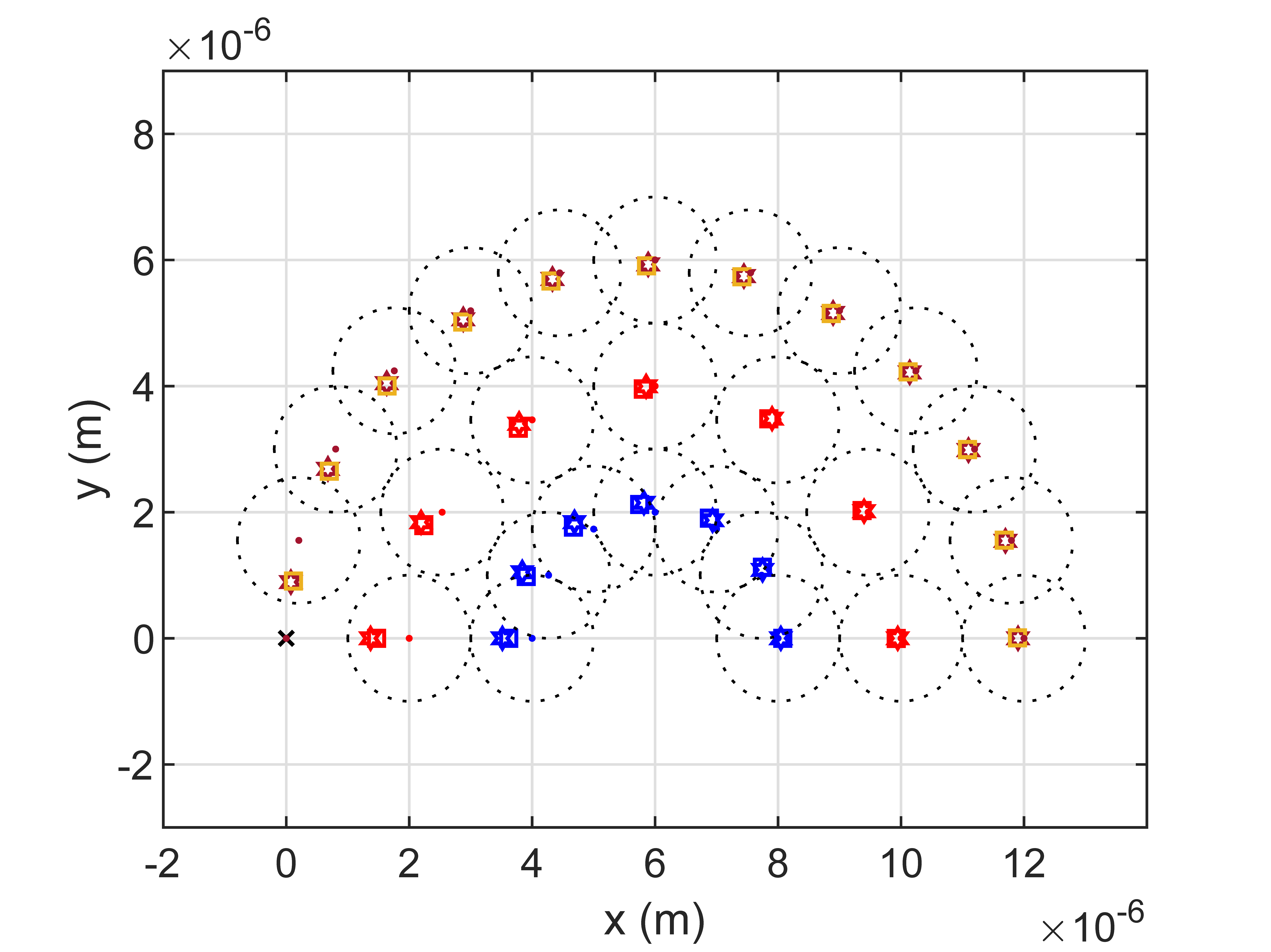}
    \caption{Interferer barycenter $B_I$ for various distances $d_{C_IC_R}$ and angles $\theta$: the agreement between the $B_I$ model predictions (hexagrams) and the simulation values (squares) is excellent for all interferer positions. }
    \label{fig - B_I model}
\end{figure}
\begin{figure*}[t!]
    \centering
    \includegraphics[width=\columnwidth]{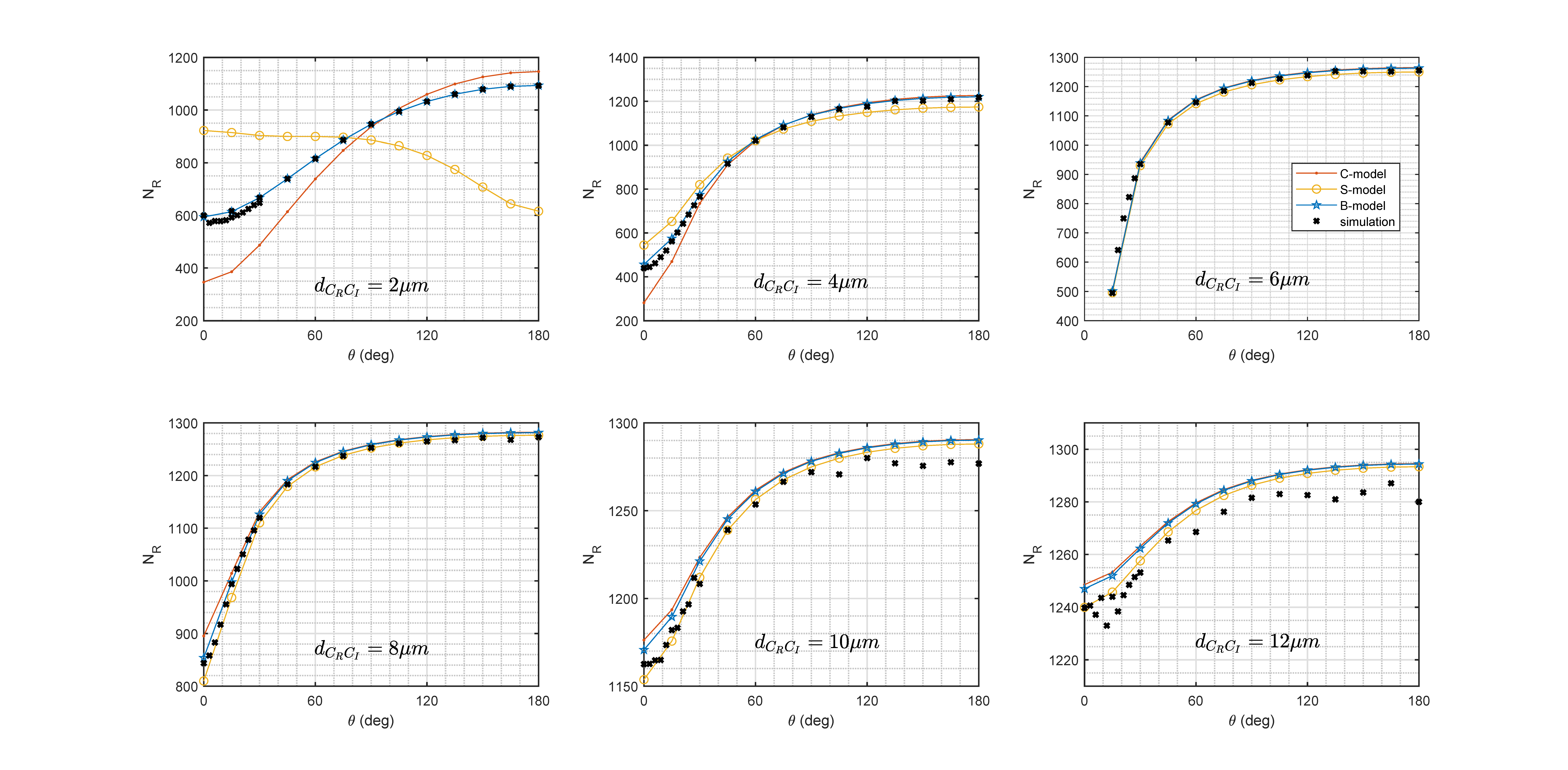}
\caption{Cumulative number of molecules $N_R(T)$ absorbed by $\mathcal{R}$ after $T=2$ s, in the scenario
     of Fig. \ref{fig:pos_cells} with $r_R=6\,\mu\mathrm{m}$, for various
     positions of $\mathcal{I}$ identified by $d_{C_RC_I}$ and $\theta$. C-model, S-model and B-model predictions versus simulation $\left( \Delta t = 10^{-6}\,\mathrm{s}\right)$.}
    \label{fig - model B, C and D}
\end{figure*}

The amount $||\boldsymbol{\delta}||$ of this
displacement is modelled as a function of the distance $d_{C_I C_R}$ between the two cells, and the assessment of this amount has been done observing the barycenter position of $\mathcal{R}$, when the interferer $\mathcal{I}$ is moved with $\theta$ from 0 to $180^\mathrm{o}$, at distances $d_{C_I C_R}~=~2,~4,~6~\mu m$. In Fig.~\ref{fig - b_R}a we
show the \textcolor{black}{position of the barycenter inside the spherical receiver drawn for the three values of $d_{C_I C_R}$. Figure~\ref{fig - b_R}b
reports with more detail the positions shown in Fig.~\ref{fig - b_R}a to better appreciate the discrepancies.} In Fig.~\ref{fig - delta} we plot the corresponding
displacement value $||\boldsymbol{\delta}||$ as a function of
$d_{C_I C_R}$ (which exhibits once again an approximately
exponential decay), together with an empirically tuned exponential
law, \textit{i.e.},
\begin{equation}
  ||\boldsymbol{\delta}\left(C_R,C_I\right)|| = 0.21 R \exp \left( \frac{-0.8(d_{C_I C_R}-2R)}{2 R} \right).
\label{eq: deltaExponentialLaw}
\end{equation}

Finally, the barycenters $B_R$ and $B_I$, according
to the proposed model are localized at 
\begin{equation}
    \begin{split}
  B_R =& \gamma \left(r_R\right) S_R + (1-\gamma \left(r_R\right)) C_R + \boldsymbol{\delta}\left(C_R,C_I\right),
   \\
  B_I =& \gamma \left(r_I\right) S_I + (1-\gamma \left(r_I\right)) C_I + \boldsymbol{\delta}\left(C_I,C_R\right).
    \end{split}
    \label{eq:barycentermodelling}
\end{equation}


In Fig.~\ref{fig - B_I model} for each drawn position of the cell
$\mathcal{I}$, we mark the barycenter computed by simulation (square) and the outcome of the barycenter model $B_I$
(hexagon) predicted by \eqref{eq:barycentermodelling}. Figure~\ref{fig - b_R} shows the same points for $\mathcal{R}$.

\subsection{Modified molecules absorption model}

The barycenter analysis and modelling allow a further refinement
of the channel impulse response model, namely B-model, in presence
of an interferer. Again, as for the S-model refinement, the
distance between the two cells is simply replaced by the
barycenter-to-center distances of the cells. 
The
system \eqref{eq - C-model equations} is modified in the following
way:
\begin{equation}
    \begin{cases}
        n_R\left(t\right) = N_T f \left( r_R,t \right) - n_I\left(t\right) \star f \left( d_{B_I C_R},t \right)
        \\
        n_I\left(t\right) = N_T f \left( r_I,t \right) - n_R\left(t\right) \star f \left( d_{B_R C_I},t \right)
    \end{cases}.
    \label{eq - model D equations}
\end{equation}

\textcolor{black}{Solution of~\eqref{eq - model D equations} is obtained as follows, (see Appendix~\ref{Ap:proof1G}) 
\begin{align}
     \resizebox{.9\hsize}{!}{$\begin{aligned}
    N_R(t) &= \frac{N_{T}R}{r_R}\sum_{n=0}^{\infty}\frac{R^{2n}}{(d_{C_{R}B_{I}}d_{C_{I}B_{R}})^{n}} \mathrm{erfc}\left(\frac{(r_R-R)+n(d_{C_{R}B_{I}}+d_{C_{I}B_{R}}-2R)}{2\sqrt{Dt}}\right)\\
    &\hspace{0.4cm}- \frac{N_{T}R^{2}}{d_{C_{R}B_{I}}r_I}\sum_{n=0}^{\infty}\frac{R^{2n}}{(d_{C_{R}B_{I}}d_{C_{I}B_{R}})^{n}} \mathrm{erfc}\left(\frac{(d_{C_{R}B_{I}}+r_I-2R)+n(d_{C_{R}B_{I}}+d_{C_{I}B_{R}}-2R}{2\sqrt{Dt}}\right)
     \end{aligned}$}\label{eq:SITO_N1_detD}
 \end{align}
}

In Fig. \ref{fig - model B, C and D} we plot the predictions of
the B-model versus simulation results and models C and S. The
agreement of the predictions of the B-model with simulation is
excellent for low values of $d_{C_I C_R}$. For large values of $d_{C_I C_R}$, model B and C predictions almost agree and they seem to overestimate the number of absorbed molecules compared to simulation. 

\section{Generalization to an arbitrary number of interferers}
\label{section - multiinterf}

The generalization of the C-model and S-model is straightforward: the effect of the complete set of competitors on
each cell assumed as the receiver is just the superposition of the
(negative) signals of each negative source of molecules. These
sources are concentrated in space in points which, in both models,
are independent of the presence of other interferers. For instance,
let $C_{I_k}$ denote the $k$th interferer center, at distance
$r_{I_k}$ from $\mathcal{T}$, and $S_{I_k}$ its
S-point. 
Equations \eqref{eq - C-model equations} are generalized into
\begin{equation}
    \begin{cases}
        n_R\left( t \right) =& N_T f \left( r_R,t \right) - \sum_k n_{I^{(k)}}\left( t \right) \star f \left( d_{C_RC_I^{(k)}},t \right)\\
        n_{I^{(k)}} \left( t \right) =& N_T f \left( r_{I^{(k)}},t \right) - n_{R}\left( t \right) \star f \left( d_{C_RC_I^{(k)}},t \right) -\\
        &-\sum_{j \neq k} n_{I^{(j)}} \left( t \right) \star f \left( d_{C_I^{(k)}C_I^{(j)}},t \right)\\
        \end{cases}
    .    
    \label{eq - multi I C-model equations}
\end{equation}

As to the S-model \eqref{eq - S-model equations}, the same
reasoning leads to the following generalization:
\begin{equation}
    \begin{cases}
        n_R\left( t \right) =& N_T f \left( r_R,t \right) - \sum_k n_{I^{(k)}}\left( t \right) \star f \left( d_{C_RS_I^{(k)}},t \right)\\
        n_{I^{(k)}} \left( t \right) =& N_T f \left( r_{I^{(k)}},t \right) - n_{R}\left( t \right) \star f \left( d_{C_I^{(k)}S_R},t \right) -\\
        &-\sum_{j \neq k} n_{I^{(j)}} \left( t \right) \star f \left( d_{C_I^{(k)}S_I^{(j)}},t \right)\\
    \end{cases}
    .
    \label{eq - multi I S-model equations}
\end{equation}


The generalization of the B-model is only slightly less
straightforward. Each negative source is concentrated in a point
(barycenter) which depends on the reciprocal position of the
interfering cells and thus is not independent of the presence of
multiple interferers. According to the barycenter modelling
proposed in Section \ref{section - B-model} we can assume the same base position
$B_0$ given by \eqref{eq:b_0}, which is independent of the
presence of interferers, and we combine all the displacements due
to the interfering cells, generalizing
\eqref{eq:barycentermodelling} to
\begin{align}
\begin{split}
 B_R =& \gamma \left(r_R\right) S_R + (1-\gamma \left(r_R\right)) C_R + \sum_k \boldsymbol{\delta}
  \left(C_R,C_{I_k}\right)\\    
 B_{I_k} =& \gamma \left(r_{I_k}\right) S_{I_k} + (1-\gamma \left(r_{I_k}\right)) C_{I_k} + \boldsymbol{\delta}
  \left(C_{I_k},C_R\right) +\\
 &+\sum_{j \neq k} \boldsymbol{\delta}
  \left(C_{I_k},C_{I_j}\right).
  \end{split}
\label{eq - b_R multinterf}
\end{align}

Once all the barycenters $B_R$, $B_{I_k}$ are computed, the
B-model \eqref{eq - model D equations} is obviously generalized to
\begin{equation}
    \begin{cases}
        n_R\left( t \right) =& N_T f \left( r_R,t \right) - \sum_k n_{I^{(k)}}\left( t \right) \star f \left( d_{C_R B_{I^{(k)}}},t \right)\\
        n_{I^{(k)}} \left( t \right) =& N_T f \left( r_{I^{(k)}},t \right) - n_{R}\left( t \right) \star f \left( d_{C_{I^{(k)}}B_R},t \right) -\\
        &-\sum_{j \neq k} n_{I^{(j)}} \left( t \right) \star f \left( d_{C_{I^{(k)}}B_{I^{(j)}}},t \right).\\
    \end{cases}
    \label{eq - multi I B-model equations}
\end{equation}

\begin{figure}[t!]
    \centering
    \includegraphics[width=0.75\columnwidth]{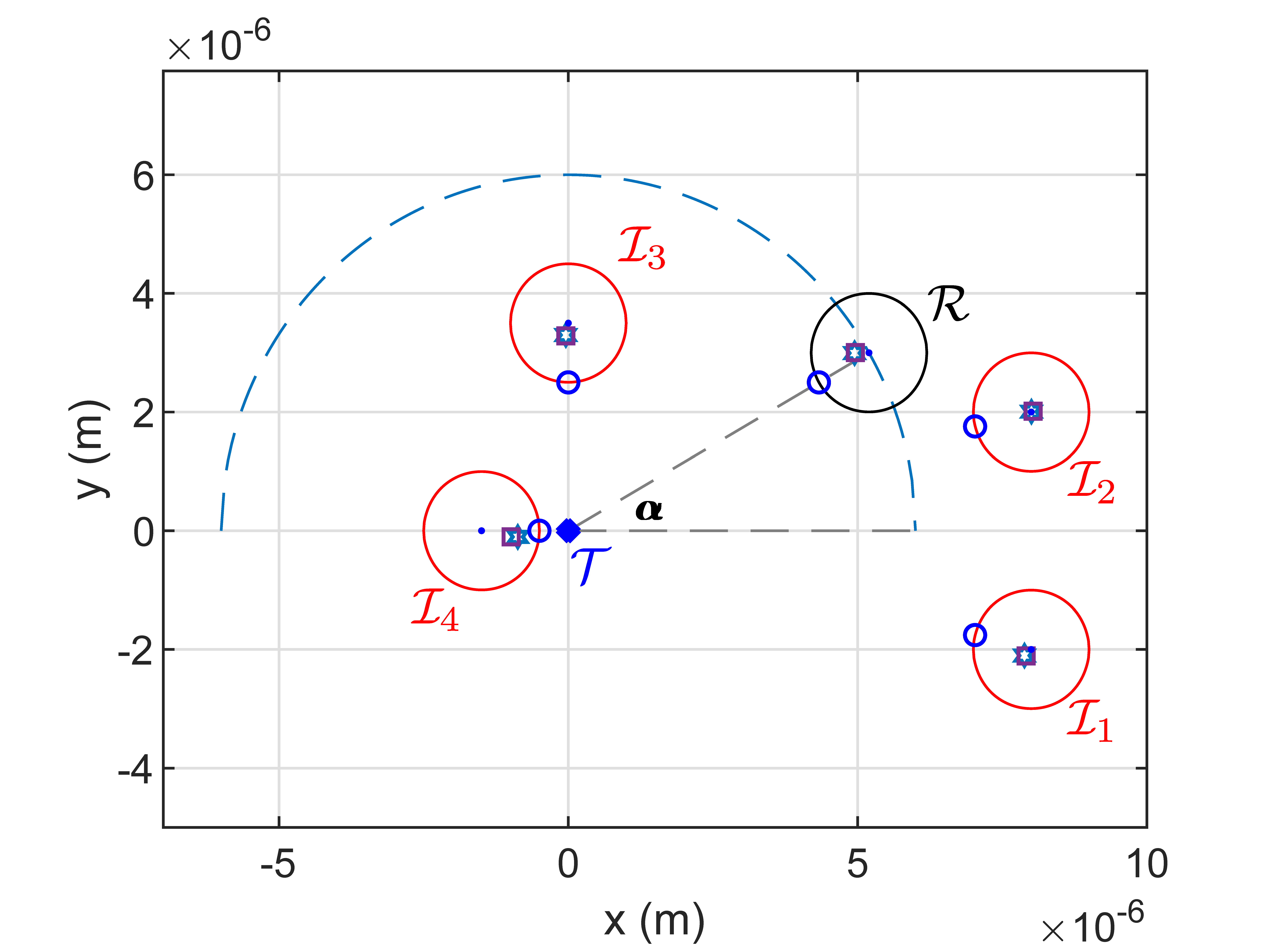}
    \caption{Example of scenario with one receiver (black line) and
    multiple interferers (red lines). The dashed line shows the path travelled by the receiver.
    Measured and modelled barycenter positions are also marked as usual (squares and hexagons, respectively).}
    \label{fig - multi-interferer scenario}
\end{figure}

\begin{figure}[t!]
    \centering
    \includegraphics[width=0.75\columnwidth]{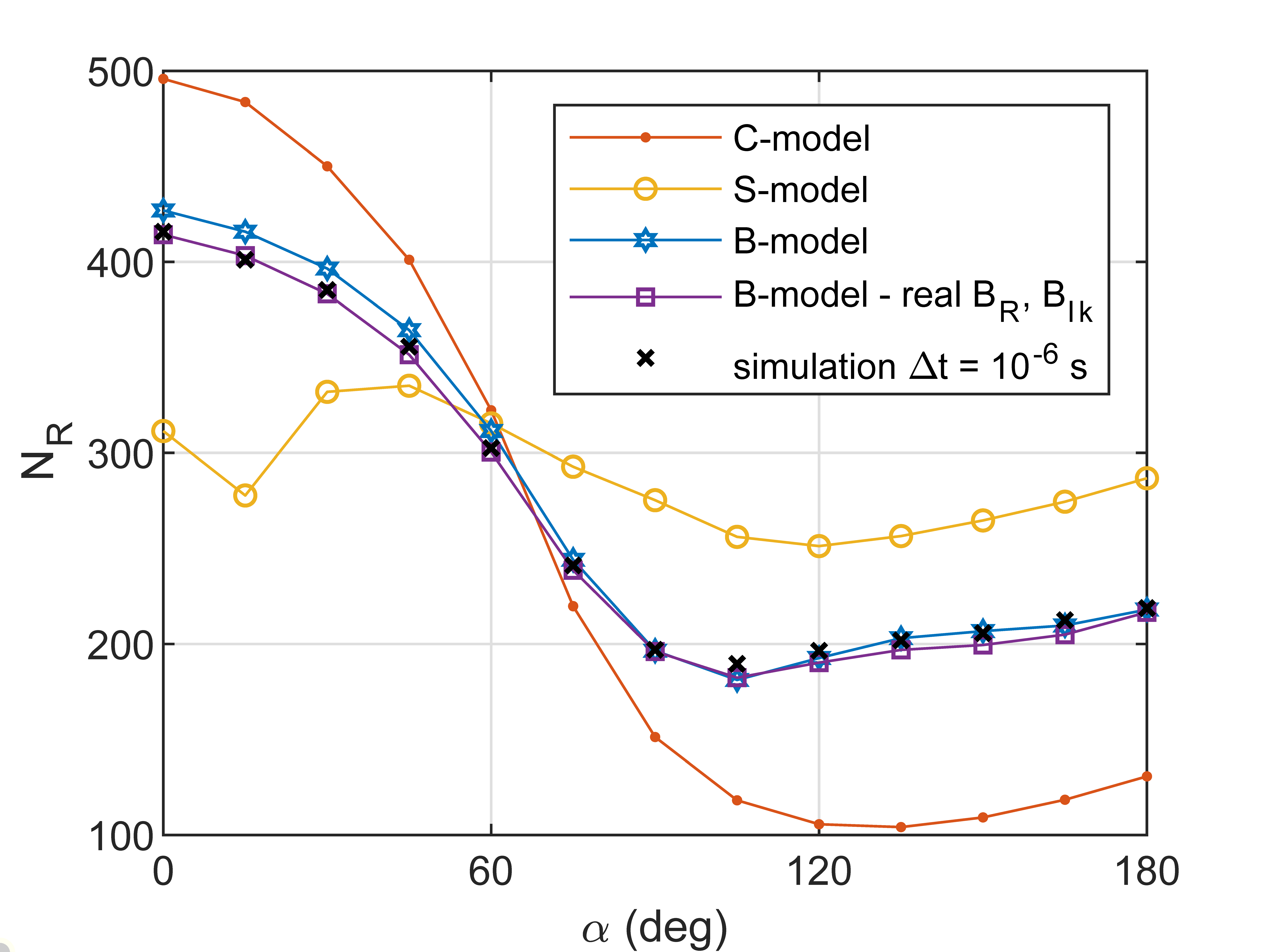}
    \caption{Cumulative number of molecules absorbed by $\mathcal{R}$ after $T=2$ s, in the scenario
     with multiple interferers of Fig. \ref{fig - multi-interferer scenario} for the various
     positions (identified by the angle $\alpha$).}
    \label{fig - multi-interferer scenario N_R(T)}
\end{figure}

\textcolor{black}{Unlike for the case of two receiving cells, we have not found a generalized closed form solution of \eqref{eq - multi I C-model equations}, \eqref{eq - multi I S-model equations} and \eqref{eq - multi I B-model equations}, yet. However, since the impulse response $f\left(d,t\right)$ is causal, we can solve the generalized systems
by numerical integration.}
To test our models generalization to the case of multiple
interferers, we consider the scenario depicted in Fig. \ref{fig -
multi-interferer scenario}. We let $\mathcal{R}$ (black) move on
the half circle described by the dashed line, experiencing the
blocking and shadowing effects of the four interfering cells
sketched with red lines. For the case drawn ($\alpha = 30^\mathrm{o}$) we also show the measured barycenters and the models from
\eqref{eq - b_R multinterf} for each cell, marked with the usual
notation (squares and hexagons, respectively).

We plot in Fig. \ref{fig - multi-interferer scenario N_R(T)} the
cumulative number $N_R(T)$ of the molecules absorbed by
$\mathcal{R}$, after $T= 2\,$s versus the angle $\alpha$. According to all models but the S-model, the shadowing effect from
$\mathcal{I}_1$ and $\mathcal{I}_2$  mildly penalizes the receiver
$\mathcal{R}$ which collects the highest number of molecules for
$\alpha<45^\mathrm{o}$. As the receiver approaches $\alpha=90^\mathrm{o}$ the number of
captured molecules quickly drops to a minimum, due to the
blocking-and-shadowing effect from $\mathcal{I}_3$. The actual
minimum is however observed for a larger angle (around $120^\mathrm{o}-130^\mathrm{o}$ according to the C-model, $100^\mathrm{o}$ for the B-model), in
an area where the combination of the blocking effect from
$\mathcal{I}_4$ on one side and $\mathcal{I}_3$ on the other
minimizes the molecules density. In fact, when $\alpha=180^\mathrm{o}$, a
mild increase in the number of absorbed molecules is observed
despite the strong blocking effect from $\mathcal{I}_4$: in that
position $\mathcal{R}$ can capture molecules from its exposed
lower side. 

Comparing the predictions to the results of simulation (marked
with crosses) we see that the B-model is much more accurate than
the C-model, and its agreement with simulation is excellent even in presence of multiple interferers, except for a light overestimation of $N_R\left(T\right)$ for $\alpha<60^\mathrm{o}$. Conversely, in such a complex
scenario, the S-model fails to correctly describe the dynamic of
the system. The reason is probably that in any point $\mathcal{R}$
is placed too close to some point $S_k$ taken as the position of
the negative source modelling the effect of $\mathcal{I}_k$. Pick
$\alpha = 15^\mathrm{o}$, for instance: $S_2$ is so close to the boundary of
$\mathcal{R}$ that model C predicts an absolute minimum, which is
not confirmed by simulation.

In Fig. \ref{fig - multi-interferer scenario N_R(T)} we also plot, marked with squares,
the curve predicted by the B-model when the real barycenter
positions, estimated by simulation, are used (B-model - real
$B_R,B_{I_k}$). In this case the agreement with simulation is excellent even for low values of $\alpha$.
Granted that there is no point in running a simulation to provide
data for the analytical-numerical model, we show this curve to
stress the fact that the B-model approach is very effective and provides
perfectly accurate predictions even in scenarios with multiple interferers,
as long as all the cells barycenters are correctly estimated. Finally, as a last comment to Fig. \ref{fig - multi-interferer scenario N_R(T)}, it can be observed that between $120^\circ$ and $180^\circ$ the estimated barycenter seems to produce more accurate result than the real one. This happens because there is a blocking effect from two interferes and to have a more accurate results from the particle-based simulation a higher time resolution would be required.

\section{Conclusion}\label{sec:conclusion}

An emerging research area in molecular communication (MC) is represented by the investigation of dynamic environments, where bio-nanomachines are described as mobile nodes in a network. While initial studies of MC have focused on static situations, where transmitting and receiving nanomachines are in fixed positions, it is now recognized that mobility allows groups of bio-nanomachines to improve collaboration and provide more complex functionalities with respect to the limited computational capabilities of a single bio-nanomachine. As a consequence, in order to evaluate the performance of mobile MC systems, there is an increasing interest to develop channel models for scenarios where there are multiple bio-nanomachines that collectively sense the molecules released by the target site.

With focus on the diffusive channel, in this paper we have studied a scenario defined by a single pointwise transmitter and multiple mobile fully-absorbing (FA) receivers. The main result is the analytical derivation of the channel impulse response between the pointwise transmitter and each of the multiple receivers that takes into account
the instantaneous relative position of the others receivers. The idea at the basis of the proposed modelling is to consider each FA receiver as a source of negative molecules and apply the superposition principle to obtain the channel response for each transmitter-receiver pair.
\textcolor{black}{As such, the channel impulse responses of all the FA receivers are linked in a system of integral equations that are numerically solved. Furthermore, for the case of two receivers a closed-form solution has been obtained as an infinite series.}
We also show that, for the case where there is only one interfering receiver that is not far from both the transmitter and/or the target receiver, the channel impulse response shape is distorted. As the most evident effect, a temporal shift of the peak in the time evolution of the molecule absorption by the receiver is observed, which is well predicted by the proposed model. The entire analysis has been validated by particle simulations and a good match is observed with the numerical solution of the derived integral equations.

\vspace{-0.8cm}
\textcolor{black}{
\appendices
\section{Analytical solution for System of equations}\label{Ap:proof1G}
As it can be seen for the case two receivers the difference between each model is only a couple of parameters in their system of equations. Hence in the following we derive solution for system of equations regardless of any specific parameter. In the end by substituting the corresponding parameter the closed-form solution of each model is achieved. The system of integral equations solved by applying Laplace transform. The following Laplace transform is well known~\cite{schiff1999laplace},
\vspace{-0.1cm}
\begin{equation}
\mathscr{L}\{\frac{a}{t^{3/2}}e^{-b/t} \}=\frac{\sqrt{\pi}}{\sqrt{b}}ae^{-2\sqrt{b}\sqrt{s}},
\end{equation}
where $\mathscr{L}$ is the Laplace transform. We can write the system of equation for the case of two receivers as
\vspace{-0.1cm}
\begin{equation}
    \begin{cases}
        n_R\left(t\right) = N_Tf \left( r_R,t \right) - n_I\left(t\right) \star f \left( d_{RI},t \right)
        \\
        n_I\left(t\right) = N_Tf \left( r_I,t \right) - n_R\left(t\right) \star f \left( d_{IR},t \right)
    \end{cases}
    \label{eq: SITO_generic_SE}
\end{equation}
where $d_{RI}$ and $d_{IR}$ can be replaced with proper parameters according to the models. 
Thus applying Laplace transform on integration of~\eqref{eq: SITO_generic_SE} results in 
\begin{equation}
    \begin{cases}\hspace{-0.05cm}
        \hat{N}_R\hspace{-0.05cm}\left(s\right)\hspace{-0.05cm} = \hspace{-0.05cm}N_T \frac{R}{s r_R}e^{-\frac{r_R\hspace{-0.02cm}-\hspace{-0.02cm}R}{\sqrt{D}}\sqrt{s}} \hspace{-0.08cm}-\hspace{-0.05cm} \hat{N}_I\hspace{-0.05cm}\left(s\right) 
        \frac{R}{d_{RI}}e^{-\frac{ d_{RI}\hspace{-0.02cm}-\hspace{-0.02cm}R}{\sqrt{D}}\sqrt{s}}
        \\\hspace{-0.05cm}
       \hat{N}_I\hspace{-0.03cm}\left(s\right) \hspace{-0.03cm} = \hspace{-0.03cm} N_T \frac{R}{sr_I}e^{-\frac{r_I\hspace{-0.02cm}-\hspace{-0.02cm}R }{\sqrt{D}}\sqrt{s}} -\hspace{-0.05cm} \hat{N}_R\hspace{-0.05cm}\left(s\right) \frac{R}{d_{IR}}e^{-\frac{\hspace{-0.02cm} d_{IR}\hspace{-0.02cm}-\hspace{-0.02cm}R \hspace{-0.02cm}}{\sqrt{D}}\sqrt{s}}
    \end{cases}\hspace{-0.3cm},
    \label{LSeq - S-model equations3}
\end{equation}
where~$\hat{N}\left(s\right) =\mathscr{L}\{N\left(t\right)\}$. Solving~\eqref{LSeq - S-model equations3} for $\hat{N}_R\hspace{-0.05cm}\left(s\right)$ gives
\begin{equation}
    \hat{N}_R(s) = N_{T}\frac{\alpha\frac{e^{-\beta\sqrt{s}}}{s}-\delta\frac{e^{-\epsilon\sqrt{s}}}{s}}{1-\kappa e^{-\gamma\sqrt{s}}},\label{eq:N1solG} 
\end{equation}
where the parameters are $\alpha \hspace{-.1cm}=\hspace{-.1cm} R/r_R$, $\beta \hspace{-.1cm}=\hspace{-.1cm} \left(r_R\hspace{-.1cm}-\hspace{-.1cm}R\right)/\sqrt{D}$, $ \delta \hspace{-.1cm}= \hspace{-.1cm}R^2/d_{RI}r_I$, $\epsilon \hspace{-.1cm} =\hspace{-.1cm} \left(d_{RI}\hspace{-.1cm}+\hspace{-.1cm}r_I\hspace{-.1cm}-\hspace{-.1cm}2R\right)/\sqrt{D}$, $\kappa = R^2/\left(d_{RI}d_{IR}\right)$, and $\gamma = \left(d_{RI}+d_{IR}-2R\right)/\sqrt{D}$.
Assuming that $|\kappa|<1$ we can replace the denominator of~\eqref{eq:N1solG} with its power expansion series
 \begin{align}
     \resizebox{.75\hsize}{!}{$\begin{aligned}
    \hat{N}_R(s) &= N_{T}\alpha\frac{e^{-\beta\sqrt{s}}}{s}\sum_{n=0}^{\infty}\kappa^{n}e^{-n\gamma\sqrt{s}}-N_{T}\delta\frac{e^{-\epsilon\sqrt{s}}}{s}\sum_{n=0}^{\infty}\kappa^{n}e^{-n\gamma\sqrt{s}}\\
    &= N_{T}\alpha\sum_{n=0}^{\infty}\kappa^{n}\frac{e^{-(\beta+n\gamma)\sqrt{s}}}{s}-N_{T}\delta\sum_{n=0}^{\infty}\kappa^{n}\frac{e^{-(\epsilon+n\gamma)\sqrt{s}}}{s}
     \end{aligned}$}\label{eq:1-2G}
 \end{align}
Knowing the following inverse Laplace transform~\cite{schiff1999laplace} 
\begin{equation}
    \mathscr{L}^{-1}\{\frac{e^{-a\sqrt{s}}}{s}\} = \mathrm{erfc}\left (\frac{a}{2\sqrt{t}} \right), \label{eq:L}
\end{equation}
where $\mathscr{L}^{-1}$ stands for the inverse Laplace transform we can apply it on~\eqref{eq:1-2G}
 \begin{align}
     \resizebox{.8\hsize}{!}{$\begin{aligned}
    N_R(t) &= N_{T}\alpha\sum_{n=0}^{\infty}\kappa^{n}\mathrm{erfc}\left(\frac{\beta+n\gamma}{2\sqrt{t}}\right) - N_{T}\delta\sum_{n=0}^{\infty}\kappa^{n}\mathrm{erfc}\left(\frac{\epsilon+n\gamma}{2\sqrt{t}}\right)
     \end{aligned}$}\label{eq:1-2Gt}
 \end{align}
Substituting the desired distance between the center of the receiver and the negative source in~\eqref{eq:1-2Gt}, according to the discussed models, gives the closed-form expression that describes the number of observed molecules of a receiver in presence of an interferer at time.}
\textcolor{black}{The C-model system of equations~\eqref{eq - C-model equations} can be seen as a special case, where the final solution \eqref{eq:1-2Gt} is simplified by $d_{RI}=d_{IR}$. }

\vspace{-0.7cm}

\textcolor{black}{
\section{Proof of convergence~\eqref{eq:SITO_N1_det}}\label{Ap:proof2}
From the physicality of the problem we expect that by locating the interferer far from the receiver, receivers observation should converge to receiver observation as there was no interferer. Equivalently, we expect that~\eqref{eq:SITO_N1_det} converges to~\eqref{eq:integ}.\\
Increasing the distance between the interferer and the receiver means that~$d_{C_{I}C_{R}}\rightarrow\infty$. Without loss of generality, we fix the position of the receiver and locate interferer at infinite. This will also results in~$r_{I}\rightarrow\infty$.
\begin{align}
     \resizebox{0.9\hsize}{!}{$\begin{aligned}
    \lim_{r_{I} \to \infty} \lim_{d_{C_{R}C_{I}} \to \infty}\Bigg(& \frac{R}{r_{R}} N_{T}\sum_{n=0}^{\infty} \left(\frac{R}{d_{C_{I}C_{R}}}\right)^{2n} \mathrm{erfc}\left ( \frac{r_{R}-R  + 2n\left( d_{C_{I}C_{R}}-R \right)}{2\sqrt{Dt}} \right )\\
    & \hspace{0.4cm}- \frac{R}{r_{I}} N_{T}\sum_{n=0}^{\infty} \left(\frac{R}{d_{C_{I}C_{R}}}\right)^{2n+1} \mathrm{erfc}\left ( \frac{r_{I}-R  + \left ( 2n+1 \right ) \left( d_{C_{I}C_{R}}-R \right)}{2\sqrt{Dt}} \right )\Bigg),
    \end{aligned}$} \label{eq:SITO_N1_lim1}
 \end{align}
 Obviously by increasing the the distance of the interferer with respect to the receiver and the transmitter all terms of~\eqref{eq:SITO_N1_lim1} converges to zero, except for the first term of summations,~$n=0$.
 \begin{align}
     \resizebox{.9\hsize}{!}{$\begin{aligned}
    \lim_{r_{I} \to \infty} \lim_{d_{C_{I}C_{R}} \to \infty}& \Bigg( \frac{N_{T}R}{r_R} \mathrm{erfc}\left ( \frac{r_R-R}{2\sqrt{Dt}} \right ) - \frac{N_{T}R}{r_I}  \left(\frac{R}{d_{C_{I}C_{R}}}\right) \mathrm{erfc}\left ( \frac{\left(r_I-R \right) + \left( d_{C_{I}C_{R}}-R \right)}{2\sqrt{Dt}} \right )\Bigg)
    \end{aligned}$} \label{eq:SITO_N1_lim2}
 \end{align}
 The first term of~\eqref{eq:SITO_N1_lim2} is independent of~$r_I$ and $d_{C_{R}C_{I}}$. The second term converges to zero.
 \begin{align}
     \resizebox{.5\hsize}{!}{$\begin{aligned}
    \lim_{r_{I} \to \infty} \lim_{d_{C_{R}C_{I}} \to \infty} N_{R}\left( t\right)& =  \frac{N_{T}R}{r_R} \mathrm{erfc}\left ( \frac{r_R-R}{2\sqrt{Dt}} \right )\\
    & =  \frac{2N_{T}R}{r_R\sqrt{\pi}} \int_{\frac{r_R-R}{2\sqrt{Dt}}}^{\infty} e^{-\tau^2} d\tau
    \end{aligned}$} \label{eq:SITO_N1_lim3}
 \end{align}
Define the new integral variable as $\tau = \left(r_R-R\right)/\left(2\sqrt{Dz}\right)$
and substitute in~\eqref{eq:SITO_N1_lim3}. The integral~\eqref{eq:SITO_N1_lim3} changes to
\begin{align}
     \resizebox{.6\hsize}{!}{$\begin{aligned}
    \lim_{r_{I} \to \infty} \lim_{d_{C_{R}C_{I}} \to \infty} N_{R}\left( t\right)& =  \frac{2N_{T}R}{r_R\sqrt{\pi}} \int_{t}^{0} -\frac{r_R-R}{4\sqrt{Dz^3}} e^{-\left(\frac{r_R-R}{2\sqrt{Dz}}\right)^2}dz\\
    & =  \frac{N_{T}R}{r_R\sqrt{\pi}} \int_{0}^{t} \frac{r_R-R}{2\sqrt{Dz^3}} e^{-\left(\frac{r_R-R}{2\sqrt{Dz}}\right)^2}dz\\
    & =  \int_{0}^{t} \frac{N_{T}R\left(r_R-R\right)}{2r_R\sqrt{\pi Dz^3}} e^{-\left(\frac{r_R-R}{2\sqrt{Dz}}\right)^2}dz\\
    & =  \int_{0}^{t} N_{T} f\left(r_{R},z \right)dz
    \end{aligned}$} \label{eq:SITO_N1_lim4}
 \end{align}
which is the same as~\eqref{eq:integ} for the number of observed particles by a receiver without any interferer.}

\bibliographystyle{IEEEtran}
\bibliography{bibliografia}

\end{document}